\documentclass{emulateapj}
\slugcomment{DRAFT \today}
\shorttitle{SDSS/2MASS M star Lum./Mass Function}
\shortauthors{Covey et al.}

\begin{document}

\title{The Luminosity and Mass Functions of Low-Mass Stars in the Galactic Disk: I. The Calibration Region\footnote{Based in part on observations obtained with the Apache Point Observatory 3.5-meter telescope, which is owned and operated by the Astrophysical Research Consortium.}}

\author{Kevin~R.~Covey\altaffilmark{1}, Suzanne~L.~Hawley\altaffilmark{2}, John~J.~Bochanski\altaffilmark{2}, Andrew A. West\altaffilmark{3}, I.~Neill~Reid\altaffilmark{4}, David A. Golimowski\altaffilmark{5}, James R. A. Davenport\altaffilmark{6}, Todd Henry\altaffilmark{7}, Alan Uomoto\altaffilmark{8}}

\altaffiltext{1}{Harvard-Smithsonian Center for Astrophysics, MS-72, 60 Garden Street, Cambridge, MA 02138; kcovey@cfa.harvard.edu}
\altaffiltext{2}{University of Washington, Department of Astronomy, Box 351580, Seattle, WA 98195}
\altaffiltext{3}{Univeristy of California Berkeley, Astronomy Department, 601 Campbell Hall, Berkeley, CA 94720-3411}
\altaffiltext{4}{Space Telescope Science Institute, Baltimore, MD 21218}
\altaffiltext{5}{Department of Physics and Astronomy, The Johns Hopkins University, 3400 North Charles Street, Baltimore, MD 21218-2686}
\altaffiltext{6}{Department of Astronomy, San Diego State University, San Diego, CA 92182-1221}
\altaffiltext{7}{Department of Physics and Astronomy, Georgia State University, Atlanta, GA 30302-4106}
\altaffiltext{8}{Carnegie Observatories, Pasadena, CA, USA}


\begin{abstract}
We present measurements of the luminosity and mass functions of low-mass stars constructed from a catalog of matched Sloan Digital Sky Survey (SDSS) and 2 Micron All Sky Survey (2MASS) detections. This photometric catalog contains more than 25,000 matched SDSS and 2MASS point sources spanning $\sim$30 square degrees on the sky. We have obtained follow-up spectroscopy, complete to J=16, of more than 500 low mass dwarf candidates within a 1 square degree sub-sample, and thousands of additional dwarf candidates in the remaining 29 square degrees. This spectroscopic sample verifies that the photometric sample is complete, uncontaminated, and unbiased at the 99\% level globally, and at the 95\% level in each color range.  We use this sample to derive the luminosity and mass functions of low-mass stars over nearly a decade in mass (0.7 M$_{\odot} >$ M$_* >$ 0.1 M$_{\odot}$).  The luminosity function of the Galactic disk is statistically consistent with that measured from volume complete samples in the solar neighborhood.  We find that the {\it logarithmically binned} mass function is best fit with an M$_c$=0.29 log-normal distribution, with a 90\% confidence interval of M$_c$=0.20--0.50.  These 90\% confidence intervals correspond to {\it linearly binned} mass functions peaking between 0.27 M$_{\odot}$ and 0.12 M$_{\odot}$, where the best fit MF turns over at 0.17 M$_{\odot}$.  A power law fit to the entire mass range sampled here, however, returns a best fit of $\alpha$=1.1 (where the Salpeter slope is $\alpha$ = 2.35); a broken power law returns $\alpha$=2.04 at masses greater than log M = -0.5 (M=0.32 M$_{\odot}$), and $\alpha$=0.2 at lower masses.  These results agree well with most previous investigations, though differences in the analytic formalisms adopted to describe those mass functions, as well as the range over which the data are fit, can give the false impression of disagreement.  Given the richness of modern-day astronomical datasets, we are entering the regime whereby stronger conclusions can be drawn by comparing the actual datapoints measured in different mass functions, rather than the results of analytic analyses that impose structure on the data {\it a priori}.  Having validated this method to generate a low-mass luminosity function from matched SDSS/2MASS datasets, future studies will extend this technique to the entirety of the SDSS footprint.
\end{abstract}

\section{Introduction}

The mass function (MF) is a fundamental property 
of stellar systems, describing the number of stars 
as a function of stellar mass.  The MF is thus a statistical measure of the 
end result of the star formation process.  Succinctly 
characterizing a stellar population, the MF also informs our 
understanding of the structure and dynamical evolution of 
stellar clusters, the Milky Way and other galaxies.  


The first measurement of the MF was conducted by \citet{Salpeter1955}; 
for higher mass stars, this seminal result 
remains essentially unchanged to the present day.  
Salpeter found that the MF can be fit as a power law, 
formally expressed as:

\begin{equation}
\label{fundamental mass function}
\Psi(M) = \frac{dN}{dM} \propto M^{-\alpha}~~stars~~pc^{-3}~~M_{\odot}^{-1}
\end{equation} 

\noindent where $\alpha =$ 2.35 is known as the `Salpeter 
slope'.  Power law MFs which find a larger fraction of low-mass stars have 
larger values of alpha and are expressed as being `steeper', while mass 
functions that find a lower fraction of low-mass stars have smaller values 
of $\alpha$ and are described as `flatter'. 

Other functional forms have been suggested to characterize the MF.  
In particular, a number of investigators \citep{Miller1979,Chabrier2005a} 
present log normal MFs, which can be expressed as

\begin{equation}
\label{eq:bestfitlognormalmf}
\xi(M) = \frac{dN}{d~{\rm log}~M} = A~{\rm exp} \bigg( \frac{- ( {\rm log}~M - {\rm log}~M_c)^2}{2 \sigma^2} \bigg)
\end{equation}

Whichever functional form they adopt, most studies agree 
that the stellar MF appears to reach 
a peak at a few tenths of a solar mass.  Power law 
characterizations of the MF require a flattening from
a Salpeter slope for M $>$ 1 M$_{\odot}$ to $\alpha 
\sim$0.5--1.7 below 1~$M_{\odot}$, while \citet{Chabrier2005a}
finds a log-normal MF with a characteristic mass of M$_c$=0.2 
is required to reproduce the stellar MF for a volume complete
sample in the solar neighborhood.  We refer the
reader to comprehensive reviews by \citet{Scalo1986}, 
\citet{Reid2000}, \citet{Kroupa2002}, \citet{Chabrier2003}, 
and \citet{Corbelli2005} for a more detailed understanding of the vast literature
devoted to measuring the stellar MF.  

Several processes have been advanced to 
explain the shape of the MF; examples include 
gravitational fragmentation \citep{Klessen1998}, 
competitive accretion 
\citep{Larson1992}, truncation of mass accretion due to radiative or
dynamical feedback 
\citep{Silk1995}, star-star interactions \citep{Reipurth2001} and the primordial
distribution of clump masses within molecular clouds \citep{Padoan2002}.  
The efficiency of each mechanism is influenced by 
physical variables, such as the metallicity and magnetic field strength
of the parent molecular cloud, the local stellar density, and the intensity 
of the surrounding radiation field.  These effects may ultimately result in 
observable MF variations as a function of environment, but at present
the observed variations can be largely attributed to uncertainties arising 
from finite sample sizes and systematic differences in methodology \citep{Chabrier2005a}, 
though intriguing signs of MF variations with environment may be emerging \citep[e.g., Taurus; ][]{Luhman2004}.

Previous studies of the Galactic disk MF have been limited, however, to samples 
of a few thousand low-mass stars (with low-mass stars 
defined for the purposes of this paper as M$_{*} <$ 0.7 M$_{\odot}$), due 
to the inability to simultaneously obtain a deep and wide photometric sample. 
\citet{Hawley2002} demonstrated that Sloan Digital 
Sky Survey (SDSS) and Two Micron All Sky Survey (2MASS) photometry of low-mass stars 
shows monotonic behavior across a wide range of colors.
The accurate, multi-color catalogs produced by SDSS and 2MASS thereby allow the 
identification and characterization of millions of low-mass stars in the 
local Galactic neighborhood, enlarging previous photometric samples of field 
low-mass dwarfs by several orders of magnitude.  The statistical power of such 
a sample makes an SDSS/2MASS catalog of low-mass dwarfs a promising avenue 
for improving our measurement of the low-mass stellar luminosity and mass
functions.

To test and calibrate this technique before applying it to the entirety of the SDSS and 
2MASS databases, we have performed a combined photometric and 
spectroscopic study of nearly 30,000 stars detected in $\sim$30 square degrees of overlap
between the SDSS and 2MASS footprints.  We describe in \S \ref{merge} the construction
of this sample by matching the SDSS, 2MASS, and Guide Star Catalog 
(GSC) photometric databases. 
In \S \ref{specsec:overall} we present the spectroscopic follow-up of
more than a third of the sample.  We verify the completeness limit of the photometric sample
in \S 4, where we also analyze the spectroscopic sample to understand the level of contamination and 
bias within the photometric catalog.  We use this photometric catalog to measure the luminosity function of low-mass stars in the Galactic neighborhood 
in \S 5, derive a measurement of the low-mass stellar mass function in \S 6, and summarize our conclusions in \S 7.  An appendix to this work presents native or
transformed SDSS photometry for stars with measured trigonometric parallaxes, useful
for constraining the empirical SDSS/2MASS color-magnitude relation applied in \S 5.  

We note that a full and complete description 
of our analysis necessarily results in a rather lengthy 
paper; we therefore suggest that many readers may wish to
skip directly to \S 5 \& 6 to concentrate upon the core
scientific findings of this work, and then return to read 
\S 2, 3 \& 4 if a deeper understanding of the assembly of
the photometric sample and the effects of incompleteness, 
contamination and bias is desired.

\section{A Unified Catalog of Survey Photometry }
\label{merge}

We analyzed SDSS, 2MASS, and GSC photometry for objects lying 
within a `calibration region' defined as the area between 
right ascensions ($\alpha$) of 3 and 15 degrees and 
declinations ($\delta$) of -1.25 to +1.25 degrees.  An abnormally large number 
of SDSS spectra of point sources were obtained in this region, allowing a test
of the efficiency and robustness of
the photometric sample assembled here (see \S \ref{specsec:SDSS}). We excluded 
a small area [13.1 $< \alpha <$ 13.3, $\delta <$ -1.05] 
surrounding HD5112, a bright (V = 4.77) M0 giant, which badly saturates 
SDSS photometry over a large area.  The area of the calibration 
region thus subtends 29.957 square degrees on the sky. 

\subsection{Individual Survey Datasets \label{photsec}}

\subsubsection{Selecting 2MASS Stars \label{photsec:2MASS}}

The Two Micron All Sky Survey \citep[2MASS;][]{Skrutskie2006} provides 
a homogeneous catalog in three near infrared 
filters ($JHK_s$) ranging from 1 to 2.4 $\mu$m.   We queried the 2MASS All Sky 
Data Release \citep{Cutri2003} via the GATOR 
interface\footnote{http://irsa.ipac.caltech.edu/applications/Gator/} for 
all objects detected within the calibration region, retaining only objects with:  

\begin{itemize}
\item{astrometric location within the boundaries of the calibration 
region;}
\item{high quality $J$ band photometry (jphqual flag $=$ `A' or 
jrdflag $=$ `1' or `3');}
\item{unique detections of catalog objects (use-src flag $=$ `1');}
\item{point source morphology (ext-key $=$ `null');}
\item{no association with the predicted location of a known minor planet, 
comet, planet or planetary satellite (mp\_flg $=$ 0).}
\end{itemize}

These selection criteria generate a sample of 30,449 stellar candidates 
with IR detections within the calibration region. 

\subsubsection{Selecting SDSS Stars \label{photsec:SDSS}}

The initial mission of the SDSS was to map a quarter of the night sky 
centered on the North Galactic cap, acquiring accurate photometry of 
100 million objects in 5 filters \citep{Fukugita1996,Gunn1998} and 
accumulating over 1 million spectra \citep{York2000}.  This mission 
is now complete, with over 9000 square degrees of photometry and over 
1.25 million spectra released to the public in Data Release 6 
\citep{Adelman-McCarthy2008}.  SDSS 
observations were obtained with a dedicated 2.5 meter telescope at the 
Apache Point Observatory \citep{Gunn2006}.  Photometric 
data were acquired as the telescope's 3 degree field imaged in 5 
filters ($u$,$g$,$r$,$i$,$z$) nearly simultaneously by scanning a great circle 
across the sky at approximately the sidereal rate.  Photometric data 
were reduced by a custom SDSS data processing pipeline
\citep[PHOTO ;][]{Lupton2001} with calibrations obtained from observations 
by a 20-inch photometric telescope at the same site 
\citep{Hogg2001,Smith2002,Tucker2006}.  The article describing the Early Data 
Release \citep{Stoughton2002} provides information on the central 
wavelengths and widths of the SDSS filters, while papers discussing 
the ``asinh'' magnitude system \citep{Lupton1999} and the SDSS 
standard star system \citep{Smith2002} provide further information 
on the calibration of the SDSS photometric system.  Astrometric 
precision and data quality assurance are also described by \citet{Pier2003} 
and \citet{Ivezic2004}.

We assembled a catalog of stellar SDSS sources within the 
calibration region from imaging data first made public as Data 
Release 2 \citep{Abazajian2004}.  SDSS stellar candidates satisfied the following criteria:

\begin{itemize}
\item{SDSS astrometric location within 0.005 degrees (15-20 $\arcsec$) of 
the calibration region (to allow matches for 2MASS objects whose proper 
motion may have carried them slightly out of the calibration region by 
the time of the SDSS imaging epoch);}
\item{observed during a run which passed basic data quality tests 
(GOOD flag set);}
\item{object represents the primary detection in the 
SDSS photometric database (PRIMARY flag set) to prevent multiple 
detections from overlap areas between adjacent SDSS runs;}
\item{morphological identification as a point source (TYPE = 6).}
\item{$i$ and $z$ magnitudes above survey completeness limits 
($i <$ 21.1, $z <$ 20.3);} 
\item{no photometric processing flaws to ensure accurate photometry 
(SATURATED flag not set in either $r$, $i$ or $z$, PEAKCENTER flag 
not set, NOTCHECKED flag not set, DEBLEND-NOPEAK flag not set for 
an object with $i$ psf error $>$ 0.2 magnitudes, PSF-FLUX-INTERP 
flag not set, BAD-COUNTS-ERROR flag not set, INTERP-CENTER and CR 
flags not both set);}
\end{itemize}

These selection criteria identified 76,966 high quality SDSS detections 
of stellar candidates within this region; the last set of quality cuts removed
1486 stellar candidates, or $\sim$1.9\% of the initial
stellar catalog.   

We note a subtle source of uncertainty in the SDSS $z$ band photometry 
which may be of more importance for this study than for most other 
uses of the SDSS database.  Spectral synthesis of L and T dwarfs 
(Burgasser, private communication) 
indicates that slight differences in the spectral response of the 
CCD detectors used for the $z$ band observations can introduce $z$
 magnitude variations up to 0.1 mags for early L through late T dwarfs. 
As we lack a robust characterization of this effect, and 
the vast majority of our sample is composed of K and M dwarfs for 
which this source of uncertainty is less important, we have not applied 
any z-band corrections to the survey photometry.

\subsubsection{Selecting GSC Stars \label{photsec:GSC}}

The depth of the SDSS comes at the expense of a relatively faint 
saturation limit of $i_{sat} \sim$ 14.  This fixed apparent magnitude
limit imposes a lower limit on the distance to SDSS stars with good 
photometric detections, and this distance limit is itself a function
of the absolute magnitude of the star.  Typical M0 dwarfs within 100 parsecs 
are saturated in SDSS imaging, for example, while M8 dwarfs only saturate in 
$i$ for distances less than 10 parsecs.  
The 2MASS faint limit, however, restricted our sample to objects 
within 100 parsecs for objects of type M8 and later.  Constructing
a luminosity function of stars 
detected in both SDSS and 2MASS then required that early and late M 
dwarfs be sampled from disjoint physical volumes.  As the calibration 
region is located at a Galactic latitude of -62 degrees, this could 
potentially have introduced a bias in our analysis related to the scale 
height of the Galactic disk. 

To include stars with saturated SDSS imaging in our catalog, we 
made use of GSC (version 2.2) optical photometry.  The GSC 
\citep[][]{Space-Telescope-Science-Institute2001} is 
based on scans of photographic plates from the Palomar and UK 
Schmidt telescopes and provides star/galaxy morphological 
classification from measurements of 500 million objects in 
photographic $B_J$ \citep[as defined by][ $\lambda \sim$ 4500 \AA]{Reid1991}
and $R_F$ ($\lambda \sim$ 6500 \AA) to a completeness limit of $R_F \sim$ 18.5.  

We assembled a catalog of GSC (version 2.2) objects brighter 
than a magnitude of 19.5 using the GSC data access 
page\footnote{http://www-gsss.stsci.edu/support/data\_access.htm}, which
we searched for morphological point sources 
(GSC Classification code $=$ 0).  These criteria identified 
33,612 stellar candidates with GSC detections within 
the calibration region.  

\subsection{Creating a Matched Sample \label{photsec:matching}}

To generate a comprehensive optical/near-infrared catalog of calibration 
region stars, we have merged the 2MASS, SDSS, and GSC 
stellar samples into a single matched catalog.  As 2MASS photometry 
provides our only source of infrared observations, only objects with 
2MASS detections were retained.  2MASS stars were matched to SDSS and 
GSC counterparts using a maximum matching radius of 5\arcsec.  
We required unique matches -- if an optical source was a potential 
counterpart for multiple NIR sources, only the closest association 
was preserved.

Of the 30,499 2MASS sources in the calibration region, 19,617 possessed 
both SDSS and GSC stellar counterparts.  An additional 5265 and 4608 
2MASS detections were identified with only a GSC or SDSS stellar 
counterpart, respectively.  The final 959 2MASS stellar candidates 
($\sim$ 3\% of the input 2MASS sample) were not matched to counterparts 
in either optical catalog.  Of these unmatched sources, 606 have 15.2 $< J <$ 16.3, 
and 753 have $J < 16.3$; given the $J=16.2$ completeness limit of our catalog (see \S \ref{quality:completelimit}), this indicates that the bulk
of these unmatched sources lie at the faint end of our sample, but still are 
confident detections. Visual analysis of SDSS imaging of the 753 $J < 16.3$ 
sources identified 73\% (550/753) as galaxies: these objects were unresolved 
by 2MASS imaging, but resolved by the optical surveys, and were therefore 
not contained in the SDSS and GSC stellar catalogs.  

The remaining 203 unmatched 2MASS sources brighter than $J=$ 16.3 were 
stellar sources, dominated by sources too faint for the GSC catalog 
and with sub-optimal SDSS photometry, and therefore excluded by the quality cuts described 
in \S \ref{photsec:SDSS}.  As these unmatched stellar sources represent 
a possible source of incompleteness in our final sample, we explore 
their properties in more detail in \S \ref{quality:completelimit}.

\subsection{Recalibrating GSC Photometry \label{photsec:recalibrateGSC} }

The 19,617 sources in common between the SDSS and GSC catalogs allowed an 
assessment of zero point errors in the GSC data.  We calculated synthetic GSC magnitudes 
for SDSS sources using equations originally derived by 
\citet{Sesar2006}, modified to account for local 
zero-point offsets:

\begin{equation}
\label{synjgsc}
B_{J,syn} = g + 0.279(g-r) + 0.06
\end{equation}

\begin{equation}
\label{synfgsc}
R_{F,syn} = r - 0.209(g-r) - 0.09
\end{equation}

\begin{figure} 
\centering 
\includegraphics[height=4in]{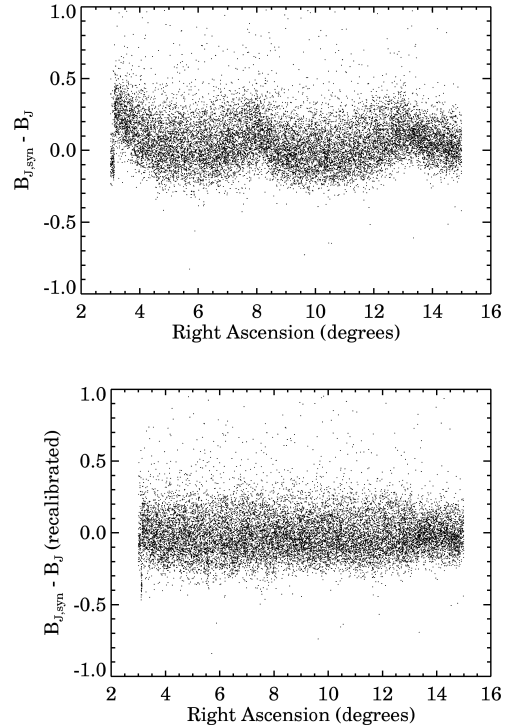} 
\caption[GSC residuals vs. RA]{Residuals of observed and synthetic 
GSC $B_J$ magnitudes calculated from SDSS photometry as a function of $\alpha$ b
efore (top) 
and after (bottom) recalibration with SDSS-based zero-points.  The structure on 
5 
degree spatial scales in the top panel suggests that GSC magnitudes, measured fr
om digitized 
POSS plates 5 degrees in size, contain systematic zero-point errors of the same 
order
as the random photometric errors.  }
\label{photfig:rawresidualsra} 
\end{figure}

The shape of the calibration region, 12 degrees wide in 
right ascension and only 2.5 degrees across in declination, implies that 
spatial gradients in imaging quality in right 
ascension are more important than gradients in declination, an expectation borne out 
in our dataset.  The top panel of Figure \ref{photfig:rawresidualsra}  
shows residuals between observed and synthetic GSC magnitudes as a function of 
right ascension.  Assuming that the digital, drift-scanned SDSS data 
has a uniform zero point across the calibration region, these residuals 
reflect changes in photographic sensitivity across POSS plates.
This expectation is supported by sharp discontinuities in the 
residuals on 5 degree spatial scales, the size of the POSS 
plates from which GSC magnitudes are measured.

To correct for spatial variations in the GSC zero point, we 
calculated the mean offset between synthetic and observed GSC 
magnitudes in 0.25 degree bins of right ascension. Applying this offset 
to all objects in that spatial bin produces recalibrated $B_J$ and 
$R_F$ magnitudes, corrected for the presence of the known zero-point
error.  Though these systematic zeropoint offsets are 
largely a function of right ascension, we have also repeated this process 
using 0.1 degrees bins in declination.  The result of this 
recalibration is illustrated in the bottom panel of Figure \ref{photfig:rawresidualsra}, which 
shows that large scale spatial gradients in the residuals have been removed.    

We adopt characteristic errors for the recalibrated GSC magnitudes based 
on the remaining residuals between the recalibrated and 
synthetic GSC magnitudes.  Gaussian errors with $\sigma=$0.13 accurately 
describe the recalibrated $B_J$ residuals across the entirety 
of the calibration region. The recalibrated $R_F$ residuals are 
well described by a Gaussian distribution with $\sigma=$ 0.1 
magnitudes for  $\alpha >$ 8.2; at $ \alpha <$ 8.2, $\sigma \sim$0.15 is required 
to fit the bulk of the sources, and a tail of sources with large errors 
is still present.  We thus conservatively assign errors of 0.2 magnitudes 
to recalibrated $B_J$ magnitudes, and 0.15 magnitudes to 
recalibrated $R_F$ magnitudes.  Given the increased 
uncertainty of the $R_F$ magnitudes for $\alpha <$ 8.2 degrees, 
we rely largely on 
$B_J$ magnitudes in conducting our analysis.

\subsection{Synthetic SDSS/2MASS Colors for GSC Sources \label{photsec:synthsdss}} 

Measuring the luminosity and mass functions from photometric observations
of the Galactic field requires the adoption of a color-magnitude relation
to estimate the absolute magnitude, and ultimately mass, of each star.
We used $i-J$ for this purpose, as \citet{Hawley2002} have demonstrated
that this color is a monotonic indicator of spectral type, and therefore 
temperature and luminosity, for late type dwarfs.  

\begin{figure} 
\centering 
\includegraphics[height=3in]{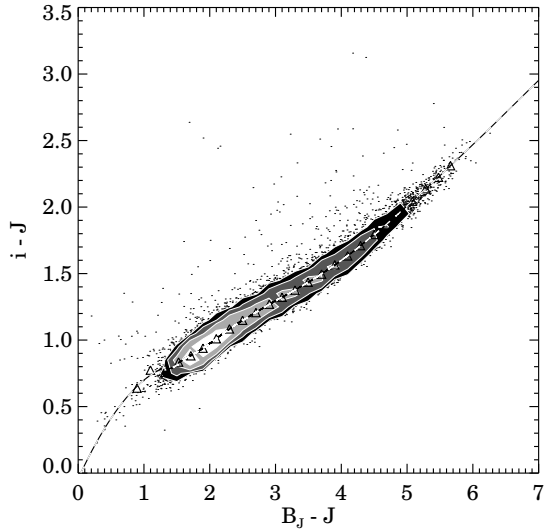} 
\caption[The $B_J-J$ vs. $i-J$ relation]{$B_J-J$ and $i-J$ 
colors for objects observed in both optical catalogs shown as black 
points and contours.  The median $i-J$ colors measured for 0.1 magnitude 
bins of $B_J-J$ color are shown as triangles; the dashed line shows 
the function used to assign synthetic $i-J$ colors as a function of 
$B_J-J$ color to sources that lack SDSS data (e.g., saturated).} 
\label{photfig:jjij} 
\end{figure}

To create a sample of uniform $i-J$ colors for the 
full photometric sample, we derived synthetic $i-J$ colors for 
objects with observed $B_J-J$ colors.  Figure \ref{photfig:jjij} 
shows the $B_J-J$ and $i-J$ colors for sources detected in 
both optical surveys.  Interpolating the $B_J-J$ colors of
our sample onto the median trend of $i-J$ as a function of 
$B_J-J$, we assigned each GSC/2MASS object that lacked bona fide 
SDSS data a synthetic $i-J$ color.  We estimated errors by 
combining in quadrature photometric uncertainties in $B_J-J$ 
and the dispersion of the $B_J-J$ vs. $i-J$ relation.  

Using the same methodology, we also derived synthetic $i-z$ 
and $r-i$ colors for all GSC sources with $B_J-J$ colors.  
For GSC sources with only $R_F$ magnitudes, we calculated the 
full suite of synthetic SDSS colors ($i-J$, $r-i$, $i-z$) using 
the same technique as applied to observed $R_F-J$ colors.  This 
minimizes the impact of the poorly recalibrated $R_F$ photometry 
on our sample: only 232 objects are detected only in $R_F$; all 
other sources have either native SDSS colors or synthetic colors 
generated from an observed $B_J-J$ color.

\section{Spectroscopic Sample}
\label{specsec:overall}

We assembled an 
extensive database of follow-up spectroscopy to test 
for contamination (e.g., background giants, mis-identified QSOs, etc.) and 
bias (e.g., systemic misclassification of numerous M2 dwarfs as rarer M6 
dwarfs due to dim, noisy photometry) in our photometrically selected sample of 
low-mass stars.

\subsection{SDSS Spectroscopy\label{specsec:SDSS}}

SDSS spectroscopic observations were made with twin fiber-fed 
spectrographs, covering wavelengths from 3800\AA\ to 9200\AA\ 
with a spectral resolution 
of $\lambda/\Delta\lambda$ $\sim$ 1800. Each 
fiber was 3\arcsec~in diameter and plugged into a 
pre-drilled metal plate allowing observations across the 3 degree field of 
view of the SDSS telescope.  A single plate accomodated 640 fibers, with 
320 going to each spectrograph.  The SDSS spectroscopic pipeline performed  
automated data reduction, producing flux calibrated spectra 
corrected for telluric absorption.

Each autumn, when the North Galactic Cap was completely inaccessable from 
Apache Point Observatory, SDSS-I focused its attention on an 
area along the celestial equator known as the Southern Equatorial Stripe,
or Stripe 82.  The observational projects conducted in Stripe 82 
with the SDSS instruments during the autumn observing seasons included 
repeat imaging to allow investigations of SDSS data quality and to conduct 
time variability studies, as well as stand-alone observing programs proposed 
by collaboration scientists which had different targeting algorithms and data 
quality benchmarks than standard survey operations.  Data products generated by these 
observing programs were released as part of DR4, and are described 
in more detail by \citet{Adelman-McCarthy2006}.  

These additional datasets motivated us to place our calibration region
in Stripe 82, as they contain thousands of stars with spectroscopic observations.  
Particularly useful to us were the SLoMaSS and `Spectra of Everything' samples, described in 
full by \citet{Bochanski2007} and \citet{Vanden-Berk2005}.
We identified stars in our sample with SDSS spectra using
a 2\arcsec~matching 
radius between the position of each spectral fiber 
and the 2MASS position of each photometric object.
Due largely to the two observing programs mentioned above, 
10,784 stars in the calibration region, or more than 
a third of our matched sample, have an SDSS spectrum. 

\subsection{Complete Observations}

For two areas within the calibration region, covering 
one square degree in total, we acquired complete 
spectroscopic samples of low-mass stellar candidates.  We 
refer to these areas as `complete region 1' (4.19 $< \alpha <$ 5.32, 
-1.04 $< \delta <$ -0.58) and `complete region 2' 
(11.68 $< \alpha <$ 12.31, -0.24 $< \delta <$ +0.8), and
the union of the two as the `main complete' sample.  
The `main complete' samples include candidate 
late type dwarfs satisfying the following criteria:

\begin{itemize}
\item{2MASS astrometric positions within the boundaries 
of complete region 1 or 2;}

\item{2MASS photometry meeting the criteria expressed in \S 
\ref{photsec:2MASS}, and $J <$ 16;}

\item{SDSS or GSC counterparts meeting the 
selection criteria outlined in \S \ref{photsec:SDSS} or \ref{photsec:GSC};}

\item{Identification as a candidate late-type dwarf by one 
of the following two color-color cuts:

\begin{itemize}

\item{SDSS counterparts satisfying an $r-i > 0.6 -1.9(i-z)$ 
color-cut, encompassing the typical colors of late-type dwarfs 
identified by \cite{Hawley2002};}

\item{Objects lacking a high-quality SDSS counterpart but identified 
as candidate late-type dwarfs with a synthetic $i-J >$ 1.2 color cut.
These candidates were only targeted in complete 
region 2, where the GSC photometry underpinning the synthetic $i-J$ color
is most reliable.}

\end{itemize}}

\end{itemize}

These cuts identified 536 targets for spectroscopic follow-up from 
an initial sample of 978 2MASS objects within the complete regions.  
The bulk of these objects (503) had SDSS counterparts meeting the 
criteria outlined in \S \ref{photsec:SDSS} for inclusion in this 
sample; a smaller number (33) had photometric flaws 
in SDSS, and we therefore resorted to synthethic, GSC-based magnitudes 
to characterize them.  

To increase the statistics of this sample at the reddest 
colors, we expanded our criteria to cover larger areas of the sky, 
targeting candidates meeting the criteria for inclusion in 
`extended red' and `super red' samples.  Candidates in these samples 
met the following criteria:

\begin{itemize}
\item{2MASS photometry meeting the criteria expressed in 
\S \ref{photsec:2MASS};}
\item{Color and astrometric cuts:
\begin{itemize}
\item{the `extended red' sample contains objects with 2MASS 
positions within -1.25 $< \delta$  $<$ 1.25 and 4 $< \alpha$  $<$ 6.5 or 
11.5 $< \alpha$  $<$ 14, and $i-z >$ 0.9, 
$i-J >$ 2.4 and $J <$ 15.8.}
\item{the `super red' sample contains objects with 2MASS positions
within the area defined by 3 $< \alpha$  $<$ 15 and
-1.25 $< \delta$  $<$ 1.25, and $i-J >$ 2.8 and $J <$ 16.}
\end{itemize}}
\end{itemize}

The union of the `main complete', `extended red', and `super red' 
samples is a set of 672 candidate dwarf stars, of which 131 
required spectroscopic observations with the ARC 3.5m telescope at APO.  Table \ref{tab:selectcomplete} 
documents the construction of each of these samples, listing the number of 2MASS sources
that fell within the boundaries of each complete sample, the number that matched to SDSS and
GSC sources, along with the subset that met the criteria for spectroscopic
observation, and finally the number of 2MASS point sources with no point source 
counterpart in either SDSS or GSC.  Table \ref{tab:completespecs} breaks down the sources of the spectra 
in each sub-sample.
 
\begin{deluxetable*}{l || c || cc | cc | c }
\tablewidth{0pt}
\tabletypesize{\scriptsize}
\tablecaption{Selecting the Complete Samples \label{tab:selectcomplete} }
\tablehead{ 
  \colhead{Sample} &  
  \colhead{2MASS sources} &  
  \multicolumn{2}{c |}{SDSS counterpart} &  
  \multicolumn{2}{c |}{GSC counterpart} &  
  \colhead{No optical} \\
  \colhead{} &  
  \colhead{in footprint} & 
  \colhead{Total} &  
  \colhead{Spectra} & 
  \colhead{Total} & 
  \colhead{Spectra} &  
  \colhead{counterpart} }
\startdata
Main Complete & 978        & 765         &  503        & 201        &  33          &  12          \\
Extended Red  & 9654       & 7382        &  101        & 2182       &  3           &  90          \\
Super Red     & 20499      & 15957       &  32         & 4246       &  0           &  296         \\
\enddata
\end{deluxetable*}

\begin{deluxetable}{l | cccr}
\tablewidth{0pt}
\tabletypesize{\scriptsize}
\tablecaption{Spectroscopic Observations \label{tab:completespecs}}
\tablehead{ 
  \colhead{Sample} & 
  \colhead{SLoMaSS} & 
  \colhead{Misc. SDSS} & 
  \colhead{APO} & 
  \colhead{Visually} \\
  \colhead{Sample} & 
  \colhead{Spectra} & 
  \colhead{Spectra} & 
  \colhead{Spectra} & 
  \colhead{Rejected} } 
\startdata
Main Complete & 421    & 15                 & 100         & 0 \\
Extended Red  & 73     & 12                 & 19          & 0 \\
Super Red     & 7      & 10                 & 12          & 3 \\
Other         & 7354   & 2892               & 60          & 0 \\
\hline
TOTAL         & 7855   & 2929               & 191         & 3 
\enddata
\end{deluxetable}

Finally, we note that SDSS spectra exist for many stars in Stripe 82 
outside the boundaries of the complete regions, and 60 ARC 3.5m spectra were also
obtained prior to the definition of the boundaries of the complete regions.  We
include these incompletely (though still densely) sampled 
sources in Table \ref{tab:completespecs} under the 
`Other' category, and used them to provide additional constraints
on the quality of the photometric sample.

Using the Dual Imaging Spectrograph (DIS) on the ARC 3.5 meter telescope 
at Apache Point Observatory, we obtained spectra of 191 candidate 
late type dwarfs in the calibration region that lacked SDSS spectra.  
The DIS spectrograph simultaneously and independently records spectra 
at blue and red 
optical wavelengths using a dichroic beamsplitter centered at 5550 \AA.  
Our observations used the `low' blue and 
`medium' red gratings and a 1.5 arcsecond (3 pixel) slit, providing 
wavelength coverage from 3800 \AA~~to 8700 \AA~~with a typical spectral 
resolution of $\sim$ 700 in the blue and $\sim$ 1000 in the red.  
Each DIS spectrum was reduced with a reduction script written in 
Pyraf, the Python-based command language for the Image Reduction and 
Analysis Facility\footnote{PyRAF is a product of the Space Telescope 
Science Institute, which is operated by AURA for NASA. IRAF is 
distributed by the National Optical Astronomy Observatories, which 
are operated by the Association of Universities for Research in 
Astronomy, Inc., under cooperative agreement with the National Science 
Foundation.}.  All spectra were 
trimmed, overscan and bias corrected, cleaned of cosmic rays, flat 
fielded, extracted, dispersion corrected, and flux calibrated using 
standard IRAF tasks.  

\section{Quality Checks \label{QualityChecks}}

Measuring the luminosity and mass functions of 
low-mass stars from a purely photometric catalog 
requires a sample of the utmost quality.  In particular, past
efforts have often been waylaid by the effects of incompleteness 
(failing to detect all stars within the adopted completeness limit), 
contamination (incorporating objects other than main sequence 
late-type dwarfs in the photometrically selected sample), and bias 
(color-dependent errors in detection efficiency or source classification).
We used our photometric and spectroscopic observations to
constrain the impact of these effects on our sample.  

\subsection{Completeness \label{quality:completelimit}}

Incompleteness in our sample would lead to a systematic underestimate 
of the true physical density of stars in the Galaxy.  Given the relative 
depths of the 2MASS and SDSS surveys, and the optical/NIR colors 
typical of low-mass stars, the inclusion of stars in our final 
sample depends primarily on the 2MASS sensitivity at faint 
magnitudes.  This expectation is borne out by the results of the 
catalog matching described in \S \ref{photsec:matching}; the vast majority 
( $> 99\%$) of 2MASS detections match to an optical 
counterpart.  The few which do not match, however, represent a 
second possible source of incompleteness in our final matched sample.  
Below we describe tests of the completeness of our sample, performed by 
empirically deriving the faint limit of 2MASS in the calibration region and
studying the population of unmatched NIR sources.  

\subsubsection{The Completeness Limit of the 2MASS survey \label{photsec:completefaint}}

\indent The canonical 2MASS 99\% completeness limit is $J = 15.9$ 
\citep{2MASSexplanatorystatement}, applicable over the entirety of the 
2MASS catalog except in areas where source confusion is the primary cause
of non-detections.  As the calibration region probes high ( $|b| > 50$) 
Galactic latitudes, stellar density is low enough that observational sensitivity
is the main source of incompleteness.
 Variations in seeing, airglow intensity, and system 
zeropoints, however, result in individual 2MASS tiles possessing true 
completeness limits up to a magnitude fainter than the canonical value.
Deeper SDSS observations, however, allow a direct measurement of the 
magnitude at which 2MASS becomes incomplete.  

To test the completeness limit of 2MASS in the calibration region, 
added an estimate of each star's $i-J$ color, derived from a least 
squares quadratic fit to the median $i-J$ vs. $i-z$ relation 
for sources with both SDSS and 2MASS detections, to their measured 
$i$ magnitudes, producing a synthetic $J$ magnitude ($J_{syn}$) for
each SDSS star. Comparing $J_{syn}$ to the actual $J$ magnitude observed
from bona fide 2MASS detections revealed that $J_{syn}$ is accurate to 
within $\sim$ 0.1 magnitudes (median error = 0.05 magnitudes, $\sigma = 
$0.13 magnitudes).

\begin{figure} 
\centering 
\includegraphics[height=3.5in]{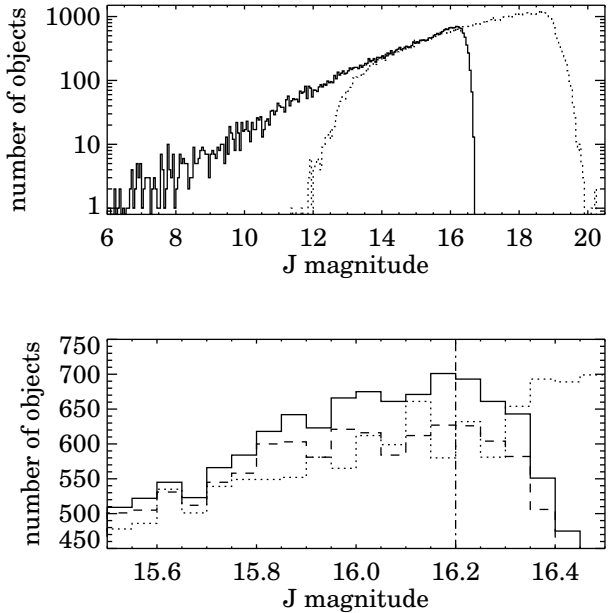} 
\caption[Examining 2MASS completeness limits]{Top: The number 
of 2MASS sources detected in the calibration region (solid black line) 
vs. $J$ magnitude compared to the number predicted on the basis of SDSS detectio
ns 
(dotted line).  Bottom: As above, focusing on the faint limit of the 
2MASS survey.  Indicated are the adopted $J=$16.2 2MASS completeness 
limit (dot-dashed line), and the number of 2MASS sources matched to optical coun
terparts 
(dashed line).  The presence of faint 2MASS `orphans' is revealed by 
the gap between the solid and dashed lines; these sources are 
discussed in \S \ref{photsec:unmatched}.} 
\label{photfig:completeness} 
\end{figure}

Figure \ref{photfig:completeness} compares the distribution of
$J$ magnitudes measured in the matched 2MASS/SDSS sample with the 
distribution of $J_{syn}$ values estimated for all SDSS sources in 
the calibration region, demonstrating that the matched 2MASS/SDSS sample 
becomes incomplete at $J =$ 16.3.  Given the $\sim$ 0.1 mag. uncertainties in 
$J_{syn}$, we select $J=$ 16.2 as a conservative estimate of the 2MASS 
completeness limit in the calibration region.  

Applying a $J <$16.2 completeness limit to our catalog reduced the 
number of 2MASS stellar candidates to 26,585.  Of these candidates, 
20,869 matched to SDSS counterparts and 23,099 matched to GSC 
counterparts, with 18,012 candidates matched to counterparts in both 
catalouges.  

\subsubsection{Understanding Unmatched 2MASS sources \label{photsec:unmatched}}

\indent The 753 2MASS sources in the calibration region with J$ <$ 16.3 and
no optical counterpart represent a second potential 
source of incompleteness.  
To understand the nature of these unmatched sources, we visually 
inspected their SDSS and 2MASS imaging.
As noted previously in \S \ref{photsec:matching}, 550 were galaxies; these objects correctly had no 
counterpart in our optical catalogs of point sources.
Of the remaining 203, only 11 appear to be spurious 2MASS detections:
nine due to the mistaken identification of a bright star's diffraction 
spike as an independent photometric object, and two due to identifying 
random sky noise fluctuations as objects.  

The remaining 192 objects appear to be genuine detections of point sources,
but whose SDSS counterpart does not meet the quality cuts discussed in 
\S \ref{photsec:SDSS}.  Table \ref{tab:flags} summarizes the 
photometric flaws preventing 2MASS sources from matching to optical 
counterparts in the SDSS catalog.  We exhaustively investigated these
unmatched stars, and provide a brief summary of the most relevant details below.  Curious readers can find a full description of this inquiry presented in 
\citet{Covey2006a}.

\begin{deluxetable}{lc}
\tablewidth{0pt}
\tabletypesize{\small}
\tablecaption{SDSS Flags For Counterparts of Unmatched 2MASS Sources\label{tab:flags}}
\tablehead{
\colhead{Stars Affected} & 
\colhead{Photometric Flaw} }
\startdata

73 & SATURATED \\
53 & PSF-FLUX-INTERP \\
22 & dubious GALAXY classification \\ 
17 & NOTCHECKED \\
13 & not PRIMARY \\
8 & PEAKCENTER \\
3 & INTERPCENTER \\
2 & undetected by PHOTO \\
1 & satellite track \\
\hline
9 & 2MASS diffraction spike \\
2 & 2MASS false detections \\
\enddata
\end{deluxetable}

SDSS counterparts for 73 of the 192 unmatched stars were eliminated 
from our sample due to the SATURATED flag being set; counter-intuitively, these
were typically faint stars, but located within the PSF wings of a 
saturated star.  The bulk of the remaining unmatched 2MASS sources 
possessed SDSS counterparts with substandard imaging, either due to 
random noise (cosmic rays) or instrumental effects 
(landing on a bad CCD column, etc).  
These flaws, indicated by the PSF-FLUX-INTERP, NOTCHECKED, PEAKCENTER, 
and INTERPCENTER flags, accounted for another 81 of the 192 2MASS stars 
unable to match to their SDSS counterpart.  The remaining 38 sources were 
excluded either 
because PHOTO identified them as extended objects (often affecting members 
of a marginally resolved visual binary), because they were 
subject to extremely rare photometric errors (sources eluding 
identification by PHOTO, or contamination of the star's PSF by the path 
of a satellite).

The Guide Star Catalog, an independent set of optical measurements,  should
protect against the loss of stars from the sample 
due to SDSS photometric errors.  Sub-standard DR4 photometry 
indicates that the majority of these sources (114/186) are too dim 
($r <$ 18.5) to be properly detected in the Guide Star Catalog.
The remaining 72 sources are typically bona fide bright stars or  
close binaries with a marginally saturated component -- the saturation 
contaminates both SDSS detections, while the pair is unresolved in GSC 
imaging.  

The tendency for unmatched 2MASS sources to possess faint and red 
counterparts in substandard SDSS photometry suggests our matching 
algorithm may be biased against the latest type stars.  To quantify 
this effect, we compared $i-J$ colors for sources with substandard 
SDSS detections to those of the full SDSS/2MASS matched sample.  
Sources with substandard SDSS detections are 
skewed towards redder $i-J$ colors than the full SDSS/2MASS sample, 
but represent less than 6\% of all sources 
for all but the reddest colors ($i-J \sim$ 2.6, where small number statistics [n$_{stars}$/bin=3] dominate).

From this analysis, we conclude that the sample is 
more than 99\% complete to $J=16.2$, and more than 94\% complete 
for every color where $i-J <$ 2.6.  Photometric flaws in 
SDSS imaging marginally affect the completeness of faint sources 
without GSC counterparts; coupled with the NIR magnitude limit 
of this sample, this effect may introduce a small color-dependent 
bias into our sample, such that we underestimate the true density of the
reddest stars at the 20\% level. 

\subsection{Spectroscopic Quality Tests}

Using the spectroscopic catalog described in Section \ref{specsec:overall}, 
we placed empirical limits on the contamination of our 
photometric catalog by objects other than low-mass dwarf stars.  
Additionally, we tested for bias within our sample, 
ensuring that photometric colors accurately predict 
spectral types (and thus luminosities 
and masses) of late type dwarfs in our catalog.  Unresolved binary 
systems represent a particularly important source of bias for
this study, and as such we discuss them separately in \S \ref{sec:lf}. 

\subsubsection{Identifying Exotic Contaminants}

We began our spectroscopic analysis by assigning spectral types to each 
spectrum, using the `Hammer' spectral typing software.  
This set of IDL routines uses 28 spectral indices to
estimate the spectral type of an input spectrum, and then allows the user
to modify the assigned type via visual comparison to a grid of dwarf 
standards.  The automated spectral typing algorithm has been tested 
with template spectra degraded to S/N $\sim 4$, returning results
accurate to $\pm$ 2 subclasses.  Tests of spectral types 
interactively assigned by multiple users agree to within one subclass.  
For a full description of the Hammer algorithm, we refer
readers to appendix A of \citet{Covey2007}\footnote{The Hammer has been made 
available for community use and can be downloaded from 
\url{http://www.cfa.harvard.edu/$\sim$kcovey/thehammer}, and/or from
the tarfile provided by \citet{Covey2007} to the \textit{AJ} to be archived with 
the electronic edition of their article}. 

Using the Hammer, we assigned spectral types to the vast majority of the 
spectra in our sample.  Of the 669 spectra in the Main Complete, Extended Red,
and Super Red spectroscopic samples, only one object could not be 
confidently assigned a stellar spectral type; it contained spectral
features from both a white dwarf and an M dwarf component.  These 
unresolved white dwarf-M dwarf (WDMD) binary systems 
have proven to be relatively 
abundant in the SDSS spectroscopic database \citep{Silvestri2006}. 

Similarly, of the 9649 sparsely sampled spectra with counterparts brighter than $J$ = 16.2, 
all but 24 were assigned stellar spectral types.  
Of the 24 non-stellar spectra, three were too noisy to confidently 
estimate a spectral type, and seven revealed additional WDMD pairs.  
The remaining 14 spectra revealed more 
exotic contaminants: four carbon stars, six unresolved galaxy populations 
(i.e, galactic bulges or clusters), and four QSOs.  The QSOs, as well 
as the seven WDMD pairs, fall within the $u-g <$ 1 and 
$g-r <$ 1.1 region that contains more than 95\% of the QSOs in the 
SDSS Quasar Catalog \citep{Schneider2005}.  The galactic 
populations and carbon stars lie within the stellar locus, unidentifiable 
by photometry alone.  Extending our analysis to the 1326 additional 
spectra fainter than $J$ = 16.2 revealed 11 more objects without 
stellar spectral types.  The majority (nine) of these objects were QSOs, 
while two objects were simply too noisy to confidently assign a spectral type.

In total, analysis of the 10,975 spectra within our sample revealed 35 
objects which cannot be assigned a stellar spectral type, implying a global
contamination rate of 0.3\% for our matched sample.  More than half (20/35)
of these exotic contaminants, however, can be eliminated by restricting 
our sample to objects outside the $u-g <$ 1 and $g-r <$ 1.1 region.  
Similarly, \citet{Ivezic2002} identified a $g-i$ vs. $i-K_s$ 
color-color cut which can also help to distinguish stars from QSOs.

\subsubsection{$i-J_{2MASS}$ as a Spectral Type Indicator}

\begin{figure} 
 \centering 
 \includegraphics[height=2in]{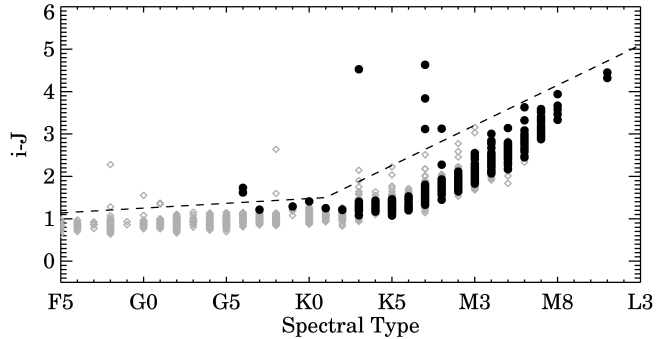} 
 \caption[Spectral types vs. $i-J$ color]{Assigned spectral types 
as a function of photometric $i-J$ color (either directly measured 
from SDSS/2MASS photometry, or transformed from GSC/2MASS photometry).  
Grey diamonds denote all calibration region objects with spectra and $J <$ 16.2,
while black circles indicate members of the complete spectroscopic samples.  
The dashed line demonstrates a cut to identify sources whose photometry 
and spectral types disagree significantly.}
 \label{fig:SpecTypeAccuracy} 
\end{figure}

Figure \ref{fig:SpecTypeAccuracy} displays spectral types 
of stars in our sample as a function of $i-J$ color.  
Consistent with previous studies
\citep{Hawley2002,West2005}, we found that $i-J$ 
predicts spectral type reliably in the M and K spectral classes. The typical 
spread of $i-J$ at a given spectral type is $\sim$ 0.1 mag, though there 
is a slight color dependence; the standard deviation of $i-J$ 
increases from 0.07 mag for G5 stars to 0.12 mag for M3 stars.  

While the vast majority of the sample shows a well-behaved color vs. 
spectral type relation, there are stars with $i-J$ colors 
significantly redder than other stars of the same spectral type.  For 
our purposes, stars with late K and early M 
spectral types but $i-J$ colors typical of late M and early L 
types are of the most concern.  With only a small number 
of bona-fide late M and early L type objects in our sample, 
consistently misclassifying even a small fraction 
of the earlier type stars could significantly inflate the 
luminosity and mass functions at the lowest masses.

To investigate the cause of these color/spectral type discrepancies, 
we inspected 20 sources with $i-J$ colors significantly redder than other 
stars of the same spectral type (this cut is shown as a dashed line in 
Figure \ref{fig:SpecTypeAccuracy}).  Four of these sources were 
early type F and G stars that saturated the SDSS, and their 
anomalous $i-J$ colors, more typical of late K/early M stars, are likely due to the
larger errors of synthetic $i-J$ colors calculated from GSC photometry.  
Given the large number of 
bona fide late-K/early-M objects in our sample, and the low incidence of
mis-classified saturated SDSS stars (4/10,940 stars in the complete 
spectroscopic sample), this effect will have a negligible
impact on our analysis.

The remaining 16 sources with anomalous $i-J$ colors are 
due to difficulties in properly associating SDSS and 2MASS sources 
into a matched catalog.  One such mismatch is SMSS (Sloan M Star Survey) J003716.5+000106.4; 
with the PSF-FLUX-INTERP flag set in the SDSS detection of this object, our 
algorithm incorrectly identifies a nearby faint star as 
this object's SDSS counterpart, resulting in a non-physical 
matched detection with an extremely red $i-J$ color.
Due to the differing spatial resolutions of
the two surveys, visual binaries with small ($\le 1.5\arcsec$) 
separations were often resolved into distinct objects by SDSS, but not
by 2MASS.  This scenario also results in an anamolously red $i-J$ color, as 
the $i$ band flux is being derived from a single component of 
the system, while the $J$ band flux is the sum of both components.

\begin{figure} 
 \centering 
 \includegraphics[height=2in]{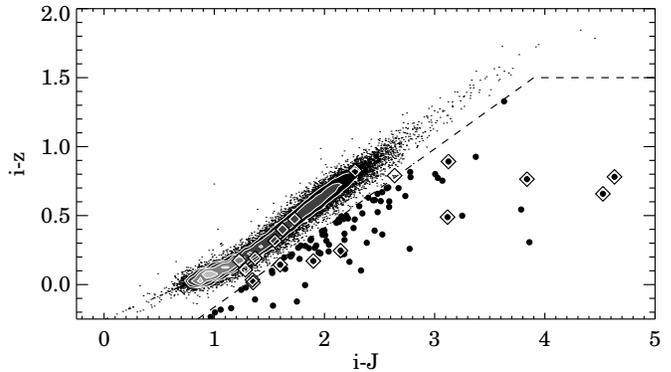} 
 \caption[Identifying mimatches in the $i-z$ vs. $i-J$ color-color 
diagram]{Mismatched SDSS and 2MASS detections in $i-z$ vs. $i-J$ 
color-color space.  Black points 
and greyscale contours show the colors of the entire photometric 
sample.  Sources identified in the full spectroscopic sample 
as possessing discrepant $i-J$ colors and spectral types are shown 
with nested black and white diamonds.  Sources with discrepant $i-J$ colors
and spectral types that fall in the middle of the $i-z$ vs. $i-J$ stellar locus
typically lack reliable SDSS detections, and have synthetic $i-J$ 
and $i-z$ colors estimated from GSC photometry and are thus consistent
with the stellar locus by construction.  Sources with $i-J$ colors 
redward of the dashed line are likely mismatches, and are shown as 
filled circles.}
 \label{ijcut} 
\end{figure}

Given the relative scarcity of objects with true $i-J$ colors $\ge 3$,
these mismatches could artificially double the number of objects 
in the reddest bins of our luminosity and mass functions.  We 
eliminated such mismatches from our sample, however, by requiring that 
the SDSS and SDSS/2MASS properties of each detection
are self-consistent.
Figure \ref{ijcut} shows a simple $i-z$ vs. 
$i-J$ color cut ($i-z$ $<$ $i-J$*0.574 - 0.738; $i-z$ $<$ 1.5) 
that we used to remove these mismatches from our sample.

\subsubsection{Kinematically Selected Subdwarf Candidates \label{specsec:kinematicsubdwarfs}}

\begin{figure} 
 \centering 
 \includegraphics[height=2.5in]{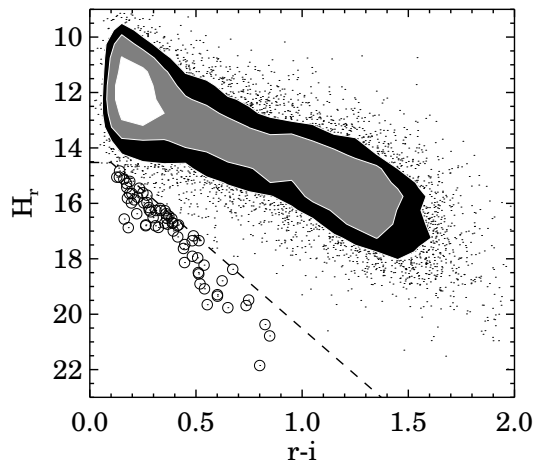} 
 \caption[Subdwarf Candidates in an SDSS reduced proper 
motion diagram]{Reduced proper motion (H$_{r}$) vs. $r-i$ 
for calibration region stars with SDSS colors and 
$\mu >$ 1\arcsec/century.  Objects with thin/thick disk 
kinematics are shown as black points and contours; 
subdwarf candidates are highlighted with open circles.  
The ($r-i$) $>$ 0.1 and H$_r >$ 13.86 + 6.666*($r-i$) 
subdwarf selection criteria is shown as a dashed line.}
 \label{reducedpropermotion} 
\end{figure}

We estimated the number of subdwarfs in our 
photometric sample by identifying objects with proper 
motions indicative of halo kinematics.  Such objects can be found 
photometrically using the reduced proper motion, 

\begin{equation}
\label{reducedpropermotion.eq}
H_r = r + 5 + 5 log \mu
\end{equation}

\noindent for a star with magnitude $r$ and proper motion $\mu$, 
expressed in arc seconds per year.  
Preliminary proper motions measured from the USNO-B 
and SDSS catalogs produce the reduced proper 
motion diagram shown in Figure \ref{reducedpropermotion} 
for objects in the calibration region with 
SDSS counterparts and $>$ 3 $\sigma$ detections of proper 
motion \citep[corresponding to a proper motion limit of 1 
\arcsec/century; ][]{Munn2004}, or more than a third of the objects in the photometric 
sample (12022/30449).  The disk population dominates 
this diagram, but a spur of objects with reduced proper motions 
lying significantly below the disk locus is clearly seen.  
We isolated candidate subdwarfs within this spur as objects 
with ($r-i$) $>$ 0.1 and H$_r >$ 13.86 + 6.666*($r-i$),
identifying 0.5\% (69/12022) of SDSS/2MASS 
objects with well-measured proper motions as candidate 
subdwarfs.  

The colors of these candidates butress their status as
subdwarf candidates.  These kinematically 
selected candidate subdwarfs are $\sim$ 0.1 mag.
bluer in $u-g$ than typical dwarfs 
with similar $g-r$ colors, consistent with the $U-B$ 
`ultraviolet excess' previously observed for subdwarfs \citep[see][Figure 2]{Bessell1979}.  
The 10 coolest ($r-i >$ 0.55) candidate subdwarfs also 
show a redward shift in $g-r$, consistent with the trend 
observed by \citet{West2004} in their sample of 
spectroscopically selected SDSS candidate subdwarfs.

We note that the 0.5\% subdwarf contamination 
rate implied by this kinematically selected sample 
overestimates the fraction of subdwarfs 
present in our sample.  The large space velocities 
of subdwarfs make them more likely than slower moving 
disk dwarfs to fulfill a 3 $\sigma$ cut on proper 
motion.  Indeed, our 0.5\% subdwarf contamination 
rate implies a local normalization of the disk and 
halo populations at the 200:1 level, a factor of  
two larger than the accepted value \citep{Reid2000}.

\subsubsection{Spectroscopically Identified Subdwarf Candidates}

M subdwarfs can 
also be identified by comparing the strength of 
metallicity sensitive spectral features, such 
as CaH and TiO \citep{Gizis1997}.  Unfortunately, no similarly 
broad features exist to identify G and K 
subdwarfs with spectra of moderate resolution 
and signal-to-noise \citep{Morell1988}.

\begin{figure} 
 \centering 
 \includegraphics[height=2in]{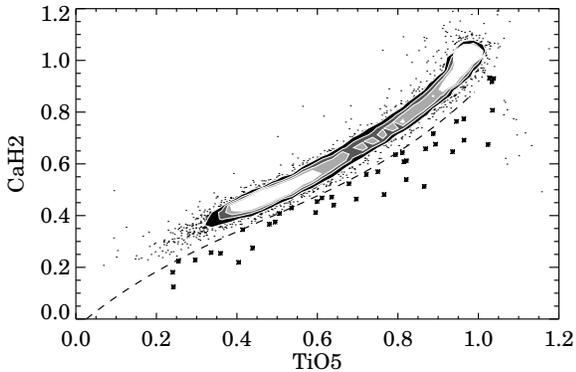} 
 \caption[Tio5 vs. CaH2 Index plot]{Measurements of the 
TiO5 and CaH2 indices defined by \citet{Reid1995}. 
Black points and grayscale 
contours show the values measured from the
subset of our sample with spectroscopic observations.  
The dashed line shows the 
boundary defined by \citet{Gizis1997} to separate 
subdwarfs from solar metallicity stars.  
Objects identified as candidate subdwarfs by this 
cut are shown as asterisks.}
 \label{tio5cah2} 
\end{figure}

We identified late-type subdwarfs in 
our sample by measuring the TiO5 and CaH2 indices 
defined by \citet{Reid1995} for our spectroscopic
sample.  The resulting TiO5 vs. CaH2 diagram is 
shown in Figure \ref{tio5cah2}, along with the 
polynomial fit defined by \citet{Gizis1997} to 
select M type subdwarfs.  Detailed abundance 
analyses by \citet{Woolf2005a} suggest that 
the subdwarf boundary defined by \citet{Gizis1997} 
identifies late type stars with [Fe/H] $\sim$ -0.5 
dex or below.  

Within the full spectroscopic sample, 
41 stars were identified as candidate subdwarfs
by their TiO5 vs. CaH2 ratio.  Visual inspection led us
to discard 22 of these sources with noisy or flawed spectra, 
leaving 19 candidate low metallicity M dwarfs.  Of 
546 M stars in the complete spectroscopic sample, 
only one was identified as a candidate late 
type subdwarf, representing an M star subdwarf 
contamination rate of 0.18\%. Similarly, of the 
7593 M stars in the full spectroscopic
 sample, the 19 candidate late type subdwarfs 
represent a 0.25\% contamination rate.  
Combining these results with the $\sim$0.5\% subdwarf fraction
implied by the kinematic analysis in \S \ref{specsec:kinematicsubdwarfs}, 
it appears that subdwarf/halo stars make up less than 0.75\%
percent of a matched SDSS/2MASS sample of late-type
stars. 

\subsubsection{Giant Stars}

Evolved stars, with effective temperatures and 
colors similar to main sequence stars, 
represent a particularly important source of 
contaminantion to a photometric catalog of main 
sequence stars.  Since giants are much more luminous
than dwarfs, a given flux limit for observational 
detection implies a larger distance limit 
for giants than for main sequence dwarfs of the same color;
effective distance limits for giants are typically 50 times 
larger, such that a survey is sensitive to giants over
a volume more than 100,000 times greater than that for 
dwarf stars. Thus, while the absolute space density of giants is low, 
they can make up a significant component of a magnitude 
limited sample of point sources.  

\citet{Bessell1988} and \citet{Worthey2006} have 
identified the $JHK_s$ and $VIK_s$ color spaces (see their
Figs. 5 and 4, respectively) 
as particularly useful for separating late-type
dwarfs and giants.  Given the size of the expected 
$JHK$ color shifts ($\sim$0.2 mag) relative to the 
accuracy of the 2MASS photometry, we can only draw 
the general conclusion that the sample is dominated by dwarf stars;
in our sample, $J-H$ becomes noticably bluer redward of $H-K_s \sim 0.15$, 
with no distinct plume of giant stars extending towards $J-H \sim$ 1 at 
$H-K_s \sim 0.15$.  Very few stars in the calibration 
region have $giK_s$ colors indicative of giants, and the tens of objects
that do make up a very small percentage
of the total number of sources detected at that $i-K_s$ color.

Spectroscopic indices allowed us to identify
G and K giant candidates within our sample, 
producing an empirical estimate of their effect 
on our analysis.  
Using 2388 spectra from numerous spectral libraries 
\citep{Torres-Dodgen1993,Danks1994,Fluks1994,
Lancon2000,Prugniel2001,Bagnulo2003,Le-Borgne2003,
Bochanski2007}, we tested spectroscopic indices 
that identify giant stars in moderate resolution, 
moderate signal to noise spectra \citep{Malyuto1997,Malyuto1999}.  
These indices 
are given in Table \ref{tab:giantindices}, with each index 
calculated as a ratio of the mean flux 
in the spectral regions identified as the index numerator and denominator.  The only exception is 
the Mg$_2$ index, which was calculated as a 
psuedo-equivalent width, consistent with its original
definition in \citet{Morrison2003}.  

\begin{deluxetable*}{lcccc}
\tablewidth{0pt}
\tabletypesize{\scriptsize}
\tablecaption{Spectral indices for Dwarf/Giant discrimination \label{tab:giantindices}}
\tablehead{
           \colhead{Band name} & 
	   \colhead{Numerator 1} & 
	   \colhead{Numerator 2} &
	   \colhead{Denominator} &
           \colhead{Reference}}
	   	  	  
\startdata

          Mg$_2$     & 4935 -- 4975 & 5303 -- 5367 & 5130 -- 5200 & \citet{Morrison2003} \\
          Mg b       & 5100 -- 5150 & 5320 -- 5420 & 5150 -- 5320 & \citet{Malyuto1997} \\
	  Na D 5900  & 5789 -- 5839 & 6020 -- 6120 & 5839 -- 6020 & \citet{Malyuto1997} \\
	  Blend 6497 & 6406 -- 6470 & 6582 -- 6637 & 6470 -- 6536 & \citet{Malyuto1999} \\
          TiO 6700   & 6425 -- 6525 & 6970 -- 7040 & 6613 -- 6832 & \citet{Malyuto1997} \\
	  CN 7900    & 7815 -- 7887 & 7970 -- 8037 & 7887 -- 7970 & \citet{Malyuto1999} \\
	  Na D 8200  & 8120 -- 8170 & 8210 -- 8260 & 8170 -- 8210 & \nodata \\
	  
	  \enddata
\end{deluxetable*}

We used Mg$_2$ vs. $g-r$ as our primary 
giant/dwarf discriminant (Morrison et al., in prep; Yanny et al., in prep.).  At red $g-r$ colors, however, the dwarf and giant 
sequences reconnect. To identify even the reddest giant stars, we 
selected giant candidates from this confused 
region of Mg$_2$ vs. $g-r$ space using three 
supplementary spectroscopic criteria 
(Na D 5900 vs. Mg b; TiO 6700 vs. Blend 6497; 
CN 7900 vs. Na D 8200) to reject likely late type dwarfs.
This technique identified 104 candidate giants from 8750 objects 
in our sample with both $g-r$ colors and spectra, representing a 
1\% global giant contamination rate. 

\begin{figure*} 
 \centering 
 \includegraphics[height=6in]{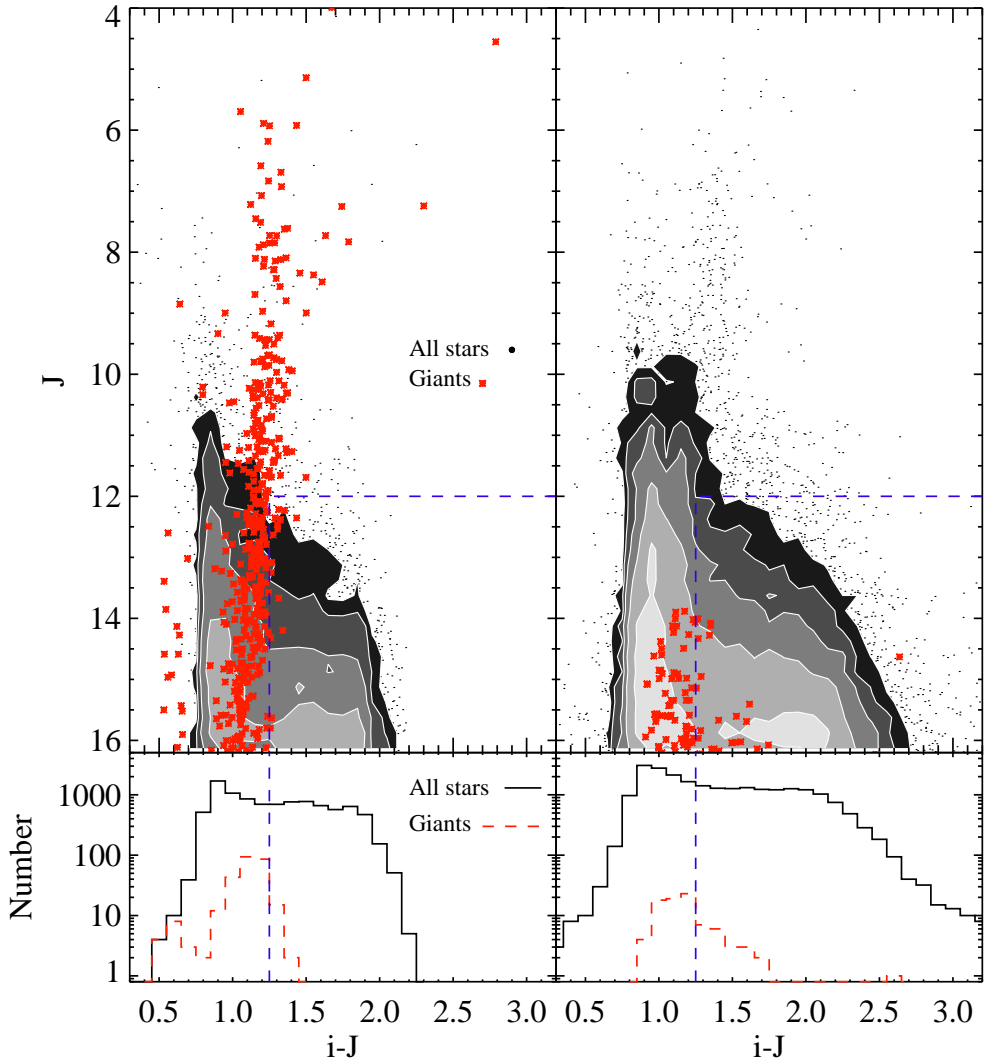} 
 \caption[$i-J$ Giant distribution]{The location of giant star 
candidates in $i-J$ vs. $J$ color-magnitude space, as predicted 
by the standard TRILEGAL galaxy model (left panels) and as 
indicated by our photometric and spectroscopic catalogs (right 
panels).  Dwarfs are shown as black points and contours in the
top two panels, while giant stars are highlighted as red asterisks and
identified using surface gravity in the TRILEGAL output or  
spectroscopic indices in the SDSS/2MASS sample.    Dashed blue lines 
show photometric cuts of $J >$ 12 and $i-J > $1.25 ($>$ K5).  Bottom panels comp
are the $i-J$ 
color distributions for all stars with $J >$ 12 (solid black line) and
for giants (dashed red lines). To reduce contamination by
giants, we restrict our luminosity function sample to sources 
with J $>$ 12 and $i-J>$ 1.25 (blue dashed lines).}
 \label{ijtrilegalcompare} 
\end{figure*}

To test if these spectroscopically identified 
candidate giants are consistent with 
standard models of the stellar population of the Milky Way,
we generated simulated SDSS/2MASS observations 
using the TRILEGAL code \citep{Girardi2005}.  
The top two panels of Figure 
\ref{ijtrilegalcompare} show the locations of giant stars in 
$i-J$ vs. $J$ color-magnitude space as predicted by the TRILEGAL
simulation using standard 
Galactic parameters and the location of the calibration region 
(top left panel) and as identified in our spectroscopic 
sample (top right panel).  The 
TRILEGAL simulation predicts that giants should reside in 
a relatively narrow locus in $i-J$ vs. $J$ color-magnitude space 
stretching from $J \sim$ 6 and $i-J \sim$ 1.3 down to 
$J \sim$ 16 and $i-J \sim$ 1.0.  
The locations of our spectroscopically 
identified candidate giant stars in 
$i-J$ vs. $J$ color-magnitude space
match these predictions well;
though the spectroscopic catalog only contains 
sources with relatively faint J magnitudes, Figure 
\ref{ijtrilegalcompare} shows the faint spectroscopically 
identified giant star candidates are indeed 
concentrated about $i-J \sim$ 1.1.  Additionally, the 
photometric sample shows a distinct plume of sources 
tracking the expected giant star locus, most prominently 
at $J \sim$ 11 and $i-J \sim$ 1.2.  

The bottom panels of Figure \ref{ijtrilegalcompare}
show histograms of the $i-J$ colors of all
stars, as well as just giants, in both the TRILEGAL
simulation and the observed data.
The TRILEGAL simulation predicts a global
contamination rate for our sample of 4\%, but this is
a strong function of color, with giant star contamination 
rising above 10\% blueward of $i-J \sim$ 1.15.  
We drastically reduced giant star contamination 
by restricting our analysis to sources with 
$J >$ 12 and $i-J >$ 1.25 ($>$ K5), bounded in Figure \ref{ijtrilegalcompare}
by blue dashed lines.  These cuts reduce giant star
contamination to a negligible level 
(less than 2\% in the bluest, most contaminated 
color bin) while sacrificing only a handful of 
bona fide low-mass dwarf stars. 

\section{The Luminosity Function}
\label{lumfunc}
The previous section demonstrates that we
can construct a matched SDSS/2MASS dataset for
stars with 12 $< J <$16.2 and 1.25 $< i-J <$ 4.5 where
incompleteness, contamination, and bias are limited to $\leq$1\% 
for the full sample, and $<$5\% for stars in a small ($\sim$ 0.1 mag) color range.
We are thus confident in the
quality of the calibration region data, and now describe the use
of this dataset to measure 
the luminosity and mass functions of the Galactic disk.

Begining with the combined photometric sample
assembled in \S \ref{merge}, we culled the catalog to ensure  
completeness  and reduce contamination 
and bias from background giants and bad matches.
In doing so, we restricted our analysis to only those 
sources meeting the following additional color and 
magnitude criteria:

\begin{itemize}
\item{completeness cut: $J < $ 16.2 \& SDSS or GSC optical counterpart (leaves 25,947 candidates);}
\item{giant contamination cut: $J > 12$ \& $i-J > $ 1.25 (leaves 13,159 candidates);}
\item{bad match cut: $i-z <$ 1.5 \& $i-z > 0.574(i-J)$ - 0.738 (leaves 13,088 candidates)} 
\item{QSO/WDMD contamination cuts\footnote{The need for $u-g$ colors to implement these cuts restricts their usage to only those objects with SDSS optical counterparts.}: $u-g >$ 1 or $g-r >$ 1.1 or $i-K < 1.5+0.8*(g-i)$ (leaves 13,064 objects in final sample).}
\end{itemize}

We will refer to this set of 13,064 objects as the Luminosity
Function (or LF) sample.  All but 
714 of these objects possess native SDSS magnitudes; the rest
possess psuedo-SDSS magnitudes transformed from GSC observations.

The calibration region probes a sight line that passes directly through the 
mid-plane of the Milky Way, where the bulk of the dust in the Galaxy is thought
to lie (h$_{dust} \sim$ 125 pc; Marshall et al. 2006).  As the vast majority of the
stars in our sample lie at distances in excess of 250 pcs, or two dust scale heights
past the midplane of the Galaxy, we assume the bulk of the extinction along
the line of sight to each star is foreground extinction.   We therefore applied 
extinction corrections to the photometry for stars in our sample, using the extinction 
estimates measured by \citet{Schlegel1998} and stored in the SDSS database (for GSC 
detections, we adopt the reddening estimate of the nearest SDSS star in 
our sample).  We use the relative extinctions in Table 6 of \citet{Schlegel1998} to apply 
corrections to our SDSS and 2MASS photometry using the R$_V$=3.1 extinction 
laws of \citet{Cardelli1989} and \citet{Odonnell1994}.  
These corrections are minimal, with a maximum E($i-J$) of 
0.11 mags and more than 95\% of the sample possessing
E($i-J$) $<$ 0.05 mags; indeed, tests indicate that neglecting
the effects of extinction produces a negligible impact on our 
results.

\subsection{The SDSS/2MASS Color-Magnitude Relation For Late-Type Stars}

Producing an LF from our photometric, 
magnitude-limited sample required us to estimate the absolute magnitude 
and distance of each star from its observed colors and an 
adopted color-magnitude relation (CMR).  The CMR for low-mass stars 
is reasonably well known in the standard Johnsons-Cousins 
photometric system; the relative youth of the SDSS, however, 
make the transformation of standard
CMRs onto the SDSS photometric system difficult
for stars with red colors. 

By spectroscopically classifying a sample of low-mass stars 
with measured SDSS and 2MASS colors, \citet{Hawley2002} derived 
photometric parallax relations for low-mass stars
in SDSS by linking distinct color-spectral type and 
absolute magnitude-spectral type relations.  Subsequent 
studies \citep{West2005,Bochanski2007,West2008} have used larger samples
to produce new measurements of the relation between 
spectral type and SDSS color, shifting the mean colors of 
a given spectral type somewhat bluer; these 
new color-spectral type relations produce new 
photometric parallax relations when 
coupled to the previously measured absolute 
magnitude-spectral type relation.  Stellar colors and magnitudes in the
SDSS and 2MASS systems have also been predicted using theoretical stellar models,
such as those calculated by \citet{Girardi2002} and \citet{Dotter2007}.

Ultimately, it would be best to determine the CMR 
by directly measuring native SDSS colors and magnitudes 
for stars with measured trigonometric parallaxes. In practice, nearly 
all such stars saturate the SDSS 2.5m camera, requiring 
observations with smaller telescopes to be 
transformed onto the 2.5m photometric system.  \citet{Williams2002} and 
\citet{Golimowski2007} have observed parallax standards 
with the 0.6 meter SDSS Photometric Telescope (PT) and the 
USNO 1 meter telescope, using the relevant 
transformations \citep{Tucker2006,Davenport2007} to 
place their photometry on the SDSS 2.5 meter system.  
We fit preliminary photometric parallax relations to describe 
the dependence of $M_J$ on $r-J$ and $i-J$ using 2MASS $JHK$ magnitudes
and transformed SDSS $ugriz$ magnitudes kindly provided to 
us by \citet{Golimowski2007} prior to publication\footnote{With an ultimate goal to 
measure the MF of low-mass stars, we chose to use CMRs 
to estimate M$_J$, for which well-established mass/luminosity relations exist 
\citep{Henry1993,Delfosse2000}.}.  These empirical CMR fits are shown in 
Figure \ref{jijCMR}, along with SDSS/2MASS 
colors and magnitudes for individual stars measured in Appendix A.

\begin{figure} 
 \centering 
 \includegraphics[height=3in]{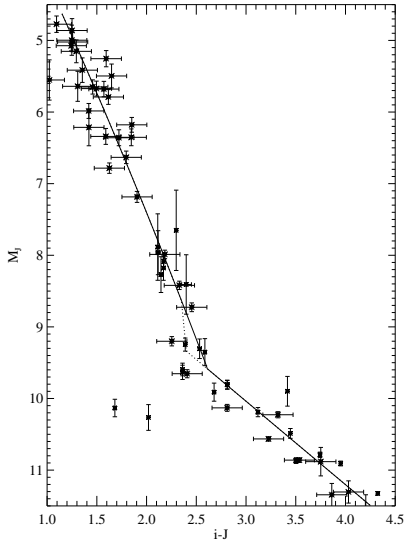} 
 \caption[Colors to Absolute Magnitudes]{Color-magnitude relations 
necessary to estimate the distribution of stellar absolute magnitudes
from the $i-J$ colors of low-mass stars in the calibration 
region.  The solid line shows a preliminary fit to the data of
\citet{Golimowski2007}, assuming a smooth fit through $i-J \sim 2.3$; 
the dotted line shows a fit to the same data, but assuming the
presence of a sharp break at $i-J \sim 2.375$.  
Shown for comparision (but not fitted directly) are the SDSS/2MASS
colors and magnitudes of individual stars as
reported in Appendix A. }
 \label{jijCMR} 
\end{figure}

The empirical CMR we have derived, however, suffers from some level of 
uncertainty.  In particular, the data are too sparsely sampled
to definitively reveal the presence or absence of a 
sharp break at $i-J \sim 2.3$, as is seen in other 
color-magnitude diagrams on the Johnson-Cousins system.  As
 discussed by \citet{Reid1997}, this feature significantly
affects the shape of the LF
inferred from a magnitude limited photometric sample.
In the absence of data clearly indicating the presence
of a sharp break in the SDSS/2MASS CMR, for our analysis we adopt the smoothly
varying CMR shown as a solid line in Fig. \ref{jijCMR}.
Understanding the important role the CMR plays in 
measuring the luminosity and mass functions of magnitude limited 
samples, however, we include uncertainty in the empirical CMR 
as a potential source of systematic error in our analysis, which 
we discuss in more detail in \S \ref{sec:lf}.

\subsection{The V$_{eff}$ Technique}

Robust space density measurements of low-mass stars 
required a careful assessment of the volume we sampled for
stars at each luminosity.  In 
particular, we needed to account for the interplay between the 
vertical density gradient of the Galactic disk and the magnitude 
limits of our catalog. We did so by using the effective volume technique.  

With roots in the $\frac{V}{V_{max}}$ technique first introduced 
by \citet{Schmidt1968}, the concept of an `effective volume' was 
first suggested by \citet{Huchra1973}, who
modified the $\frac{V}{V_{max}}$ test to accomodate a spatial gradient as a function 
of position on the sky. \citet{Felten1976} extended the V$_{eff}$ formalism to 
include gradients as a function of depth along the line of sight and proved that the
sum of $\frac{1}{V_{eff}}$ for a sample of objects can serve as an unbiased estimator 
of their spatial density, even in situations where the objects are distributed 
inhomogeneously.  

The effective volume of an object can be expressed mathematically as

\begin{equation}
\label{eq:veff}
V_{eff} = \Omega \int_{dmin}^{dmax} r^2 \frac{\rho}{\rho_0} dr
\end{equation}

\noindent where $\Omega$ is the observed solid angle, $r$ is the 
position along the line of sight to an object, $dmin$ is the minimum distance at which
the object could be detected, $dmax$ is the maximum distance at which the object could
be detected, and $\frac{\rho}{\rho_0}$ describes the ratio of the sampled density to the 
local density as a function of position along the line of sight.  
For non-uniform source distributions, $\frac{\rho}{\rho_0}$ effectively 
weights each volume element along the line of sight according to the 
expected likelihood of finding a source in that volume, such that sampling equivalent
effective volumes should result in samples with similar numbers of sources.

Using this methodology, space densities (represented by $\Phi$) 
are given by the sum of $\frac{1}{V_{eff}}$ for 
each source within the sample: 

\begin{equation}
\label{eq:nonuniformphi}
\Phi = \sum_N \frac{1}{V_{eff}} = \sum_N \frac{1}{\Omega \int_{dmin}^{dmax} r^2 \frac{\rho}{\rho_0} dr}
\end{equation}

As shown by \citet{Huchra1973}, the uncertainty in this estimator of $\Phi$ is: 

\begin{equation}
\label{eq:variancephi}
\sigma_{\Phi} = \sqrt{\sum_N \frac{1}{V_{eff}^2}}
\end{equation}

Combining the bright and faint limits of our survey sample ($J =$ 12 and 
$J =$ 16.2 respectively) with the value of $M_J$ estimated for a single source 
allows us to calculate its individual distance limits, \textit{dmin} and \textit{dmax}.
Thus, all that remains in the calculation of $V_{eff}$, and thus $\Phi(M_J)$, is to integrate 
$r^2 \frac{\rho}{\rho_0}$ between \textit{dmin} and \textit{dmax}.  

\subsection{Adopted Galactic Structure Model}

Adopting a standard multi-component Galactic model allows us to calculate the 
stellar density profile expected along a given line of sight; the total density, $\rho_{tot}$, 
is simply the sum of the density profiles of the individual disk and halo components.  
Expressed as a function of position within the Galaxy (using the standard 
Galactic coordinates R and Z), these profiles are:

\begin{eqnarray}
\label{eq:galrho}
\rho_{thin} = \exp^{\Big(-\frac{|Z|}{H_{thin}}\Big)} \;\exp^{\Big(-\frac{|R|}{L_{thin}}\Big)} \nonumber \\
\rho_{thick} = \exp^{\Big(-\frac{|Z|}{H_{thick}}\Big)} \;\exp^{\Big(-\frac{|R|}{L_{thick}}\Big)} \nonumber \\
\rho_{halo} = \Bigg( \sqrt{R^2 + (Z\times \frac{a}{c})^2} \Bigg)^{r_{halo}} \nonumber \\
\rho_{tot}  = f_{thin}\frac{\rho_{thin}}{\rho_{thin,0}} \; + \; f_{thick}\frac{\rho_{thick}}{\rho_{thick,0}} \; + \; f_{halo}\frac{\rho_{halo}}{\rho_{halo,0}} \nonumber
\end{eqnarray}

The total stellar density at a given point in the Galaxy thus depends on 
the nine parameters given in Table \ref{tab:galparams}.
Given the Galactic latitude ({\it b} $\sim$ -62) and 
the color and magnitude limits of the calibration region sample, 
$\frac{\rho}{\rho_0}$ is set primarily by the vertical 
profile of the thin disk; the thick disk only begins to 
dominate observed star counts at distances of $\sim$1.25 
kpc, where even our most intrinsically luminous stars 
approach the faint limit of our sample.

\begin{deluxetable}{llc}
\tablewidth{0pt}
\tabletypesize{\scriptsize}
\tablecaption{Galactic Structure Parameters \label{tab:galparams}}
\tablehead{
           \colhead{Parameter name} & 
	   \colhead{Parameter description} & 
	   \colhead{Adopted value} }
\startdata

$f_{thin}$ & thin disk stellar density  & (1 - $f_{thick}$ - $f_{halo}$) \\
          & in the solar neighborhood &  \\
$H_{thin}$ & thin disk scale height        & 280 pc \\
$L_{thin}$ & thin disk scale length & 2500 pc \\
$f_{thick}$ & thick disk stellar density  & 5\% \\
          & in the solar neighborhood &  \\
$H_{thick}$ & thick disk scale height         & 900 pc \\
$L_{thick}$ & thick disk scale length  & 3500 pc \\
c/a         & halo flattening parameter   & 0.7 \\
          & axial ratio c/a         & \\
$r_{halo}$ & exponent of halo power  & -2.75 \\
          & law density gradient           & \\
$f_{halo}$ & halo stellar density    & 0.15\%  \\
          & in the solar neighborhood &
\enddata
\end{deluxetable}

In principle, the parameters in Table \ref{tab:galparams} can be 
constrained by fitting observed star counts to the output 
of Galactic stellar population models produced with 
a range of Galactic structure parameters.  The best
estimates, however, are produced by comparing models to 
star counts along multiple lines of sight that sample regions
of the Galaxy dominated by different Galactic components. 
Even in this best case, there often remain significant degeneracies
between various Galactic structure parameters \citep[e.g., $f_{thick}$ 
and $H_{thick}$; see Fig. 1 of ][]{Siegel2002}.  

\begin{figure} 
 \centering 
 \includegraphics[height=2.5in]{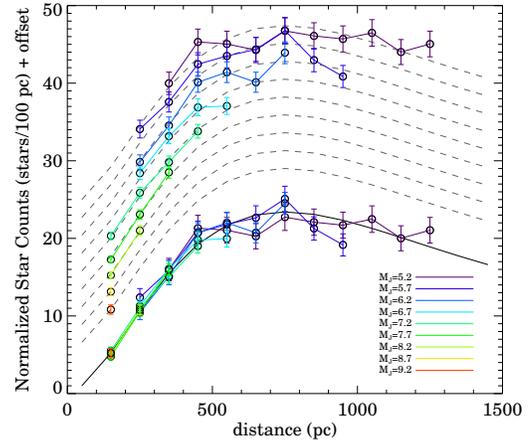} 
 \caption[Galaxy Model Vs. Observed Star Counts]{Star counts as a 
function of distance for stars with
similar M$_J$ values (colored circles). For clarity, each observed 
profile is shown twice: once compared to a common star count profile
predicted for the Galactic structure paramaters adopted here
(solid black line, bottom), and once compared to the same 
model profile but with a unique vertical offset (dashed grey line).
Observed star count profiles have been normalized such that the
total number of stars along the line of sight matches the prediction 
of the model, allowing sets of stars with different local density 
normalizations to be compared to the same model profile.  The
Galactic structure model adopted here produces a good agreement 
between the observed and predicted star count profiles.}
 \label{data-model-starcounts} 
\end{figure}

As the calibration region
probes only a single sight line through the Galaxy, attempts to
fit the Galactic model directly from the observed star counts 
produced unphysical results with significant degeneracies between
parameters.  A follow-up paper in this series (Bochanski et al. 2009, in prep.)
will present robust Galactic structure parameters measured from low-mass
stars detected across the entirety of the SDSS footprint; for now,
however, we calculated $\frac{\rho}{\rho_0}$ using standard 
Galactic structure parameters from the literature 
\citep[see Table 1 of ][]{Siegel2002, Juric2008} which
are shown in Table \ref{tab:galparams}.  As Figure
\ref{data-model-starcounts} demonstrates, the profile of 
stellar counts predicted by this model is in good 
agreement with the observed star counts in the calibration region
over the entire magnitude range of our sample.

\subsection{The Luminosity Function of the Galactic Disk}\label{sec:lf}

\begin{figure} 
 \centering 
 \includegraphics[height=4.5in]{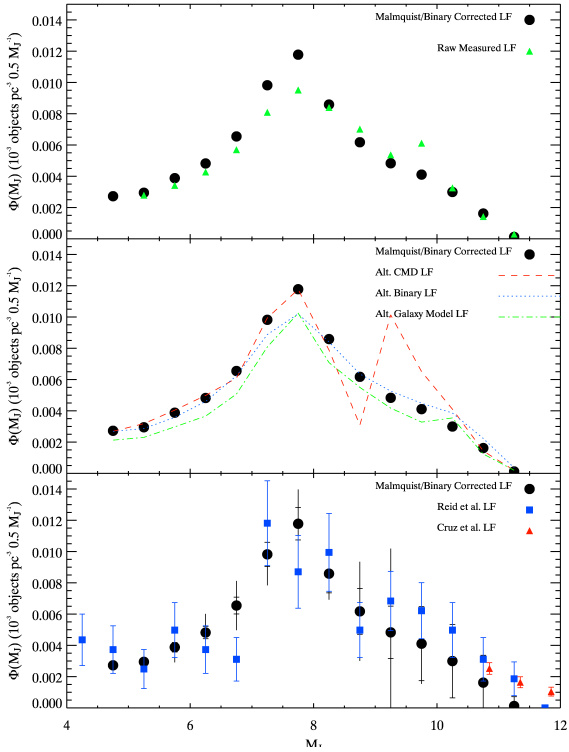} 
 \caption[Luminosity Function of the Galactic Disk]{The calibration region 
luminosity function.  \textit{Top Panel:} System luminosity function 
measured from the calibration region (green triangles), and the resultant
single-star luminosity function, after accounting for the impact of 
unresolved binaries and Malmquist type effects (black circles).  
\textit{Middle Panel:} Systematic uncertainties in the luminosity function, 
shown by comparing the calibration region single-star luminosity function 
(black circles) with results obtained when adopting a 
color-magnitude relation with a sharp break at $\sim$M4 ($i-J=$2.375; 
red dashed line), a Galactic model with a $\sim$20 
scale height (green dash-dot line), and correcting for binaries assuming 
a flat mass ratio distribution (blue dotted line).  \textit{Bottom Panel:} 
The calibration region single-star luminosity function (black circles), 
including error bars for statistical (hatted) and systematic (un-hatted) 
effects.  Shown for comparison are luminosity functions 
derived from the volume complete 8 pc sample 
\citep[blue filled squares; ][]{Reid2002} and for late-M/L dwarfs within 
20 pc \citep[red filled triangles; ][]{Cruz2007}.}
 \label{fig:lf} 
\end{figure}

Applying the V$_{eff}$ technique to the LF sample,
we obtained a first measure of the LF of 
the Galactic disk, shown in the top panel of Fig. \ref{fig:lf} as green filled diamonds.  
Derived from a magnitude-limited sample of stars with distances 
up to a kiloparsec or more, this measurement
suffers from Malmquist-type effects and represents 
the LF of \textit{systems}, not 
individual stars, in the Galactic disk.  

To account for the effects of unresolved binaries and Malmquist-like biases, 
we constructed an algorithm to reveal the intrinsic {\it stellar} LF required to reproduce the raw {\it system} LF measured
from the calibration region sample.  This routine 
uses a Monte Carlo approach to
produce a simulated photometric sample given an input 
LF, CMR, and parameters to describe the
adopted Galactic model and binary population.  We then
conduct a V$_{eff}$ analysis on this 
synthetic sample, which incorporates both observational errors
and the intrinsic width of the CMR.  Using the difference
between the synthetic `measured' LFs and the actual raw LF 
measured from the calibration region to adjust the input LF used
to run the next simulation, we iteratively converged on the true LF 
required to replicate our observed LF in the 
presence of these observational biases.  

For our primary analysis, we adopted the 
fiducial Galactic model outlined in Table \ref{tab:galparams}.  
We began by generating a synthetic set of star counts, 
randomly sampling the raw calibration region LF as an initial guess for the
input LF and accounting for the increase 
in V$_{eff}$  towards brighter M$_J$.  We used the 
smooth empirical CMR shown as a solid line in Fig. \ref{jijCMR} and the 
M$_J$ vs. mass relation derived by \citet{Delfosse2000}\footnote{We note 
that the \citet{Delfosse2000} M$_J$ vs. mass relations as published 
and archived on the A\&A website and the 
arXiv preprint server contain typographical errors.  The original authors confirm that the correct relation
is Log $\frac{M}{M_{\odot}}$ = (10$^{-3} \times (1.6 + 6.01 M_J + 14.888 M_J^2 -5.3557 M_J^3 + 0.285181 M_J^4)$. (T. Forveille, priv. communication, 2008)} to assign
masses and $i-J$ colors to each star.
We included an intrinsic dispersion in the CMR by sampling
color offsets from a normal distribution with 
$\sigma_{color} \sim 0.1$ mag, or equivalently, magnitude 
offsets with $\sigma_{mag} \sim 0.5$ mag.  

We accounted for the presence of unresolved binaries in our
sample by randomly assigning secondary companions to synthetic 
stars with a mass dependent 
binary fraction, $f_{bin}(M) = 0.45 - (0.7-M_{p})/4$, 
consistent with observations of the stellar binary fraction in the Galactic 
disk \citep{Duquennoy1991,Reid1997,Delfosse2004,Lada2006,Burgasser2007}.  
In our model, companion star luminosities were randomly selected from the same 
input LF used to generate the primary stars, but
secondaries were required to have lower luminosities than their primaries.  
This quasi-independent sampling of the LF nicely
reproduces the observed correlation between mass ratio and primary mass, 
producing a mass ratio distribution which peaks at 0.4 for stars 
more massive than 0.6 $M_{\odot}$ and at 0.9 for stars less 
massive than 0.2 $M_{\odot}$.

We assumed that all binary systems in our sample are unresolved,
based on the relatively poor spatial sampling of the 2MASS 
detectors (2\arcsec pixels).  This assumption is not
uniformly correct, as resolving a binary depends
on the ratio of this resolution limit to the binary's apparent
separation, itself a function of both the
physical separation of the companions and the distance
to the system.  Systems with an $M_J = 8$ primary, for example,
are resolved only if their separation exceeds 30-200 AU, depending 
on the location of the system within the volume to which we 
are sensitive to such stars; the situation is even worse for 
instrinsically brighter stars, which are drawn from a more distant
volume and thus have smaller apparent separations for equivalent
physical separations.  As most binary stars have separations less 
than 100 AU, and our sample is dominated by stars at the largest 
distances, we treat all simulated binary systems as unresolved. 
Fluxes from each companion are merged to produce system magnitudes
and colors, at which point random observational errors are added 
to produce a psuedo-observed photometric sample.  

We then measured the LF of this sample using the same 
algorithm as applied to our observed sample, and used the difference
between the two measured luminosity functions to modify the input LF used
to produce the synthetic sample.  After a small (n$\sim$5) number of 
iterations, the LF measured from the synthetic sample converges to that measured
from the calibration region dataset.  The final input LF whose `measured' LF 
reproduces the raw calibration region LF is therefore a measure of the 
true LF of the calibration region after correcting for unresolved 
binaries and Malmquist-type biases, as each effect is included
in producing the catalog of synthetic stars.

The calibration region LF with corrections for Malmquist biases and unresolved 
binaries is shown with black circles in each panel of Fig. \ref{fig:lf}.  
Somewhat counter-intuitively, aside from making the peak at $M_J = 7$ 
more prominent, the corrected LF is very
similar to the uncorrected LF we measure from the calibration 
region.  Applying binary and Malmquist corrections 
independently showed that the two effects offset one another
to a large degree; as discussed extensively by \citet{Stobie1989}, 
the Malmquist correction shifts the peak of the LF to 
brighter magnitudes, primarily by decreasing the
densities measured on the faint side of the LF peak.  
Accounting for unresolved binary systems, however, 
has the opposite effect -- splitting binary systems into
individual components increases the source density in the 
{\it stellar} LF above that of the {\it system} LF, and is most
important for fainter stars, which are most likely to be hidden
as secondaries to more luminous primaries.  
The primary impact of these corrections, therefore, is to increase the 
peak density of the 
LF (at M$_J = 7.5$) by 20\%; outside the 7 $< M_J <$ 8 range,
the {\it corrected stellar} LF is remarkably similar to the 
{\it measured system} LF.

To test how sensitive our LF measurement is to our
underlying assumptions, we re-derived the LF after 
modifying the adopted CMR, Galactic model, and binarity 
prescription.  We varied, in turn, the binary 
prescription to assign secondaries assuming a flat mass 
ratio distribution, the Galactic model by increasing the 
scale height of the thin disk by 50 parsecs, and the CMR 
by adopting the red dashed empirical fit in Fig. \ref{jijCMR}
that includes a sharp jump in magnitude at $i-J=2.375$. 
The LFs derived under each of these modified assumptions 
are shown in the middle panel of Fig. \ref{fig:lf}.

We characterized the systematic uncertainties in our
analysis as the fractional change induced in the LF by 
modifying each portion of the LF measurement algorithm;
adding each of these individual uncertainties in 
quadrature produced a global estimate of the 
systematic uncertainties in our analysis, which 
is shown as an additional set of error bars
for the corrected single-star LF in the bottom
panel of Fig. \ref{fig:lf}.  
The binary correction introduces a relatively small 
$\sim$5\% systematic uncertainty, rising above 10\% only at the faintest
magnitudes.  The LF is somewhat more sensitive to the Galaxy model adopted 
in the V$_{eff}$ analysis, typically producing a 20\% 
uncertainty in the derived LF.  The uncertainty in
the LF due to the Galaxy model is more pronounced at brighter 
magnitudes, where the volume sampled is several scale heights 
above the plane of the Milky Way, and thus the densities 
expected by the V$_{eff}$ calculation are more sensitive to 
the adopted thin disk scale height.  

The uncertainty in the exact 
behavior of the CMR at $i-J \sim 2.375$, however, introduces the
largest systematic uncertainty into our analysis; LFs produced
using the two empirical fits in Fig. \ref{jijCMR} differ
by more than 90\% at M$_J = $ 9.25.  Reducing this uncertainty
is clearly necessary to realize the full promise of the
SDSS and 2MASS to characterize the low-mass stellar content of the 
Galactic disk.  Doing so will require a concerted effort
to transform photometry of trigonometric parallax standards from
moderate aperature telescopes onto the native photometric system of
the SDSS 2.5m.  

The bottom panel of Figure \ref{fig:lf} displays the
single-star calibration region LF along with 
error bars accounting for both statistical and systematic effects.
Shown for comparison are the low-mass stellar LFs 
measured by \citet{Reid2002} and \citet{Cruz2007} from volume complete samples
in the solar neighborhood; with the exception of the $M_J=6.75$ bin, the 
calibration region LF agrees with the LFs measured 
from the volume complete samples, given the uncertainties in each measurement.  

\begin{deluxetable}{lcccc}
\tablewidth{0pt}
\tabletypesize{\scriptsize}
\tablecaption{The Luminosity Function of the Calibration Region\label{tab:calibregionlf}}
\tablehead{
           \colhead{} & 
	   \colhead{Raw} & 
	   \colhead{Corrected} &
           \colhead{} & 
	   \colhead{}  \\
           \colhead{M$_J$} & 
	   \colhead{LF} & 
	   \colhead{LF} &
           \colhead{$\sigma_{\rm stat}$} & 
	   \colhead{$\sigma_{\rm sys}$} }
\startdata
  5.25 & 2.79 &  2.99 & 0.25 & 0.79 \\
  5.75 & 3.41 &  3.82 & 0.31 & 0.83 \\
  6.25 & 4.27 &  4.83 & 0.39 & 1.20 \\
  6.75 & 5.70 &  6.53 & 0.54 & 1.30 \\
  7.25 & 8.09 &  9.64 & 0.77 & 1.55 \\
  7.75 & 9.52 & 11.33 & 1.04 & 1.84 \\
  8.25 & 8.40 &  8.08 & 1.23 & 0.72 \\
  8.75 & 7.01 &  6.51 & 1.46 & 2.82 \\
  9.25 & 5.37 &  4.71 & 1.67 & 4.54 \\
  9.75 & 6.12 &  4.60 & 2.37 & 1.31 \\
 10.25 & 3.26 &  3.37 & 2.35 & 0.79 \\
 10.75 & 1.42 &  1.10 & 1.75 & 1.27 \\
 11.25 & 0.30 &  0.41 & 0.59 & 0.28 \\
\enddata
\tablecomments{LFs and uncertainties reported in units of $10^{-3}$ stars $(0.5 M_J)^{-1} pc^{-3}$}
\end{deluxetable}

\section{The Mass Function}

\begin{figure} 
 \centering 
 \includegraphics[height=4.5in]{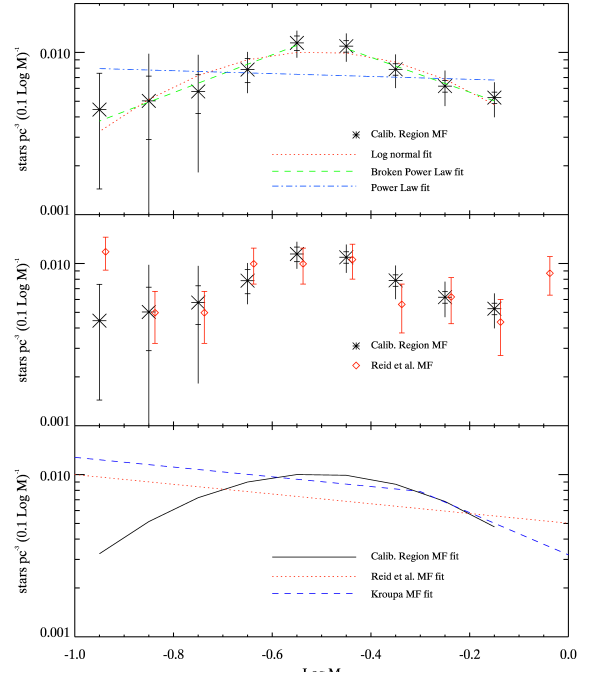} 
 \caption[]{The \textit{single-star} (not system) mass function ($\frac{dN}{d lo
g M}$) of the Galactic disk as measured from the calibration region sample (aste
risks; hatted error bars show statistical errors, unhatted error bars show syste
matic uncertainties).  \textit{Top Panel: } Analytic fits to the calibration reg
ion MF data; log normal, power law, and broken power law fits to the data are sh
own as dotted, dot-dashed, and dashed lines respectively.  An F-test indicates t
hat the log normal fit provides the greatest fidelity given the number of parame
ters used to produce the fit.  \textit{Middle Panel: } A comparison between the 
calibration region MF and that derived by \citet{Reid2002} from the 8-pc volume 
complete sample (red diamonds); the two measurements agree within the uncertaint
ies for all but the lowest mass point.   \textit{Bottom Panel: } Analytic descri
ptions of the \textit{single-star} MFs presented by \citet{Reid2002} and \citet{
Kroupa2002}, shown as red dotted and blue dashed lines respectively, compared to
 the log normal fit adopted here (solid black line).  While the data underlying 
these fits agree well, the analytic descriptions diverge significantly, particul
arly at the low-mass end.}
 \label{fig:mf} 
\end{figure}

The same simulations that allowed us to correct our observed
LF for unresolved binaries and Malmquist effects also 
provided an opportunity to measure the stellar mass function.  
Applying the $V_{eff}$ technique to the synthetic stellar sample, but 
binning by mass instead of M$_J$ magnitude, produced the MF shown 
in the top panel of Figure \ref{fig:mf}, again with both systematic 
and statistical error bars.  The {\it logarithmically binned} mass 
function of the calibration region has a clear peak at 
M$_* \sim$0.3 M$_{\odot}$.  We note that the systematic error bars, 
due mainly to the uncertainty in the color magnitude relation, dominate 
the statistical error bars for all mass bins; until the SDSS/2MASS 
color-magnitude relation is clarified, we will be unable to reap the 
full benefit of a sample of this size.

We used a Markwardt minimization
to fit various analytic forms to this data, with results shown in 
the top panel of Fig. \ref{fig:mf}.  We fit a log normal form to the 
data, finding best fit values of  A=0.1 \citep[in units of stars (Log M$_{\odot}$)$^{-1} pc^{-3}$ for comparison with the values 
reported by ][]{Chabrier2003}, M$_c$=0.29, and $\sigma$=0.32; we also fit a 
single ($\alpha=$1.1) and broken power law ($\alpha_{log M > -0.5}$=2.04; 
$\alpha_{log M < -0.5}$=0.2) to the data as well.  An F-test
evaluates the quality of an analytic fit to a dataset relative to the number
of degrees of freedom in the fit.  Applying an F-test to these fits revealed
that the log normal formalism produces a quantifiably better fit than a 
single power law: there is a less than 1\% chance that the underlying 
distribution is truly a single power-law and that the improved quality 
of the log-normal fit is a random occurance.  A dual power law fit provides 
an even better fit than either the single power law or the
log normal form, but at the cost of additional degrees of freedom; the F-test 
indicates there is a 35\% chance that the dual power law fit does not
provide a statistically superior fit to the data, and as such we adopted
the log normal fit as the most efficient analytic description of our data.

M$_c$ represents the peak of the {\it logarithmically binned}
mass function, such that while the density of stars per logarithmic
mass bin declines for masses less than M$_c$, the width of each
bin in {\it linear} mass units is also changing as well.  As a result,
a log normal MF with a given M$_c$ corresponds to a {\it linearly binned} MF
whose peak occurs at smaller masses than M$_c$.  The log normal fit 
described above corresponds to a {\it linearly binned} MF
that turns over at 0.17 M$_{\odot}$.

The MF we measure here agrees well with previous investigations.  
Using a Monte Carlo routine to refit the MF after applying random offsets consistent with each datapoint's 
total error budget, we derive 90\% confidence intervals for the log-normal
fits' parameters of A=0.08--0.12, M$_c$=0.20--0.50, and $\sigma$=0.22--0.37.  
These agree well with the values reported by 
\citet{Chabrier2005a}, with the exception of $\sigma$, for which Chabrier 
finds a somewhat larger value of 0.55, outside the bounds of our 90\% 
confidence interval.  This may result from the difference in the range 
of masses studied here and by Chabrier; Chabrier's work attempts to 
describe the MF well above and below the masses sampled 
here, which may drive the fit to larger values of $\sigma$, and wider 
MF peaks.  We also include in the middle panel of Fig. 
\ref{fig:mf} the individual datapoints for the MF constructed by 
\citet{Reid2002}, which agree within 1 $\sigma$ with the datapoints 
reported here for all but the lowest mass bin.  Similarly, the $\alpha=1.1$ power law 
fit to the entire mass range sampled here is
a good match to the $\alpha$=1.2 power law measured by \citet{Reid2002}.  

\begin{deluxetable}{lccc}
\tablewidth{0pt}
\tabletypesize{\scriptsize}
\tablecaption{The Mass Function of the Calibration Region\label{tab:calibregionmf}}
\tablehead{
           \colhead{Log M$_{*}$} & 
	   \colhead{MF $\rho$} &
           \colhead{$\sigma_{\rm stat.}$} & 
	   \colhead{$\sigma_{\rm sys.}$} }
\startdata
-0.15 &  5.42 & 0.44 & 1.29 \\
-0.25 &  6.13 & 0.50 & 1.53 \\
-0.35 &  7.91 & 0.65 & 1.52 \\
-0.45 & 10.76 & 0.89 & 1.73 \\
-0.55 & 10.82 & 1.18 & 1.54 \\
-0.65 &  7.40 & 1.28 & 1.37 \\
-0.75 &  6.07 & 1.59 & 3.54 \\
-0.85 &  4.73 & 1.93 & 3.50 \\
-0.95 &  5.61 & 3.38 & 1.47 \\
\enddata
\tablecomments{MF densities and uncertainties are reported in units of $10^{-3}$ stars $(0.1 {\rm log} M_{*})^{-1} pc^{-3}$.}
\end{deluxetable}

Simply comparing analytic fits from other studies to the MF data 
measured here, however, can give the potentially false impression 
of disagreement in the underlying MFs.  This is due in part to the 
sensitivity of analytic fits to the exact formalism chosen to represent 
the data, as well the exact range of masses spanned by the fit.  
As shown in the bottom panel of Fig. \ref{fig:mf}, the log normal description
of the MF data reported here diverges strongly from the analytic fits
reported by \citet{Reid2002} and \citet{Kroupa2002}, which would
seem to provide evidence for significant MF differences.  In 
particular, the log normal fit presented here predicts 
an MF that peaks in the stellar regime, while the break points
adopted for the power-law fits measured by \citet{Reid2002} and 
\citet{Kroupa2002} imply that the MF rises all the way to the sub-stellar 
limit.  As shown previously, however, the individual MF datapoints derived 
here and by \citet{Reid2002} agree well, revealing that these varations are 
due more to differences in the formalisms adopted to describe the 
MF than to confident detections of significant MF differences.  

Traditionally,
the utility of analytic descriptions of the data has been to produce a more
robust description of a potentially noisy dataset, collapsing the many degrees of freedom
of the sample to improve the quality of the measurement of a reduced set of parameters.
Given the richness of modern-day astronomical datasets, however, where
one can construct samples of stellar populations that contain thousands, and 
in some cases millions, of objects useful for deriving mass functions, 
the underlying data can produce MFs with vanishingly small statistical 
error bars, and systematic errors are not reduced by describing the 
data with analytic fits.  As a result, we are entering the regime where 
we create more noise than signal by comparing analytic descriptions of the MF, which impose
structure on the data {\it a priori}, rather than comparing actual MF datapoints
with statistical and systematic error bars.

Having validated this method to generate a low-mass luminosity function from matched SDSS/2MASS datasets, future studies (Bochanski et al., in prep.) will extend this technique to the entirety of the SDSS footprint, producing simultaneous measurements of the low-mass MF and Galactic structure parameters of the thin and thick disks.

\section{Conclusions}

\begin{enumerate}

\item{We have identified a sample of more than 
13,000 candidate low-mass stars detected within
30 square degrees of 2MASS, SDSS and/or GSC imaging.
Empirical tests verify the sample is 
more than 99\% complete to $J=16.2$, and more than
90\% complete at even the reddest colors.}

\item{Analysis of the spectroscopic sample reveals that exotic 
contaminants, such as QSOs, extragalactic stellar populations, CVs, 
and white dwarf/M dwarf pairs, comprise less than 0.3\% of our matched sample.}

\item{Photometric $i-J$ color provides a reliable proxy
for stellar spectral type and T$_{eff}$; we measure 
dispersions of 1 subclass for a given 
0.1 magnitude slice in $i-J$, or of 0.1 magnitude in $i-J$ 
for objects with a given MK spectral type.}

\item{In a small percentage of cases (20/9624; 0.02\%), a 
star's observed spectral type differs significantly from 
that expected on the basis of its photometric $i-J$ color.  
This usually occurs when the 2MASS and SDSS detections of 
a close visual binary are improperly matched, and we have 
identified an $i-z$ vs. $i-J$ color cut which can be used 
to eliminate such spurious detections from the sample.}

\item{Analysis of the kinematic and spectroscopic
properties of our sample reveal 88 candidate subdwarfs,
representing a subdwarf fraction of 0.68\% for our 
broader sample.  As the bulk of the candidate 
subdwarfs (69/88) were identified kinematically,
and proper motion catalogs are biased towards 
objects with large space motions such as subdwarfs,
this is an upper limit to the 
true subdwarf fraction of our photometric sample.}

\item{TRILEGAL simulations of the Galactic stellar population 
predict giant stars will contaminate our sample at
the 4\% level globally, but with giants exceeding 10\% of 
stars with $i-J \sim$ 1.15.  Candidate giant stars identified
in our sample using spectral indices verify the colors of 
giants predicted by the TRILEGAL simulation.  Restricting our 
analysis to sources with $J >$ 12 and $i-J >$ 1.25, 
however, reduce contamination of our 
sample by giant stars to a managable level 
(less than 2\% in the bluest, most contaminated 
color bin) while sacrificing only a handful of 
bona fide low-mass dwarf stars.}

\item{After correcting for unresolved binaries and
Malmquist effects, we find the luminosity function
of the Galactic disk is consistent with that measured 
from volume complete samples of the solar neighborhood.
The dominant systematic uncertainty in our analysis is
due to the remaining uncertainty in the SDSS/2MASS CMR, 
which translates into a 70\% uncertainty at the peak of
the stellar LF.  Systematic effects due to the binary correction
prescription and the adopted Galactic model are comparatively
small, at the 10-20\% level.}

\item{The logarithmically binned mass function measured from our sample
peaks at 0.3 M$_{\odot}$ and agrees within 1$\sigma$ for all but one
data point of the mass function measured by \citet{Reid2002} 
from their volume complete sample.  A log normal characterization
 with A=0.1 \citep[in units of stars (Log M$_{\odot}$)$^{-1} pc^{-3}$ for comparison with the values 
reported by ][]{Chabrier2003}, M$_c$=0.29, and $\sigma$=0.32 provides
an adequate description of the data, with 90\% confidence intervals 
for each parameter of A=0.08--0.12 ,
M$_c$=0.20--0.50, and $\sigma$=0.22--0.37. These 90\% confidence intervals correspond to {\it linearly binned} mass functions peaking between 0.27 M$_{\odot}$ and 0.12 M$_{\odot}$, where the best fit MF turns over at 0.17 M$_{\odot}$.  A power law fit to the entire mass range sampled here, however, returns a best fit of $\alpha$=1.1 (where the Salpeter slope is $\alpha$ = 2.35); a broken power law returns $\alpha$=2.04 at masses greater than log M = -0.5 (M=0.32 M$_{\odot}$), and $\alpha$=0.2 at lower masses.}

\item{We emphasize that comparisions of analytic MF fits, rather than the
underlying data, can give the false impression of MF variations
even when the underlying data agree well. We are entering the regime where 
we create more noise than signal by comparing analytic descriptions of the MF, which impose
structure on the data {\it a priori}, rather than comparing the actual MF datapoints
themselves.}

\end{enumerate}

\acknowledgements

We thank Michael Meyer and Charles Lada for helping clarify the current empirical constraints on the fraction and properties of low-mass multiple systems, August Muench and Thomas Greene for useful discussions of the mass function of low-mass stars in young clusters and the effects of extinction, and Chris McKee for a careful and thoughtful reading of a previous draft of this paper.  We also owe a debt of gratitude to Connie Rockosi for assistance in obtaining and analyzing SDSS data on occassions too numerous to count, and to Russet, Jack, Bill and John, the APO observing specialists whose assistance, patience, and occasional skirting of the telescope's strict wind limit were instrumental in helping obtain the complete spectoscopic sample necessary for this work.  We also thank the anonomyous referee, whose comments have improved this manuscript, resulting in a clearer, more concise presentation of this work.

Funding for the SDSS and SDSS-II has been provided by the Alfred P. Sloan Foundation, the Participating Institutions, the National Science Foundation, the U.S. Department of Energy, the National Aeronautics and Space Administration, the Japanese Monbukagakusho, the Max Planck Society, and the Higher Education Funding Council for England. The SDSS Web Site is http://www.sdss.org/.

The SDSS is managed by the Astrophysical Research Consortium for the Participating Institutions. The Participating Institutions are the American Museum of Natural History, Astrophysical Institute Potsdam, University of Basel, University of Cambridge, Case Western Reserve University, University of Chicago, Drexel University, Fermilab, the Institute for Advanced Study, the Japan Participation Group, Johns Hopkins University, the Joint Institute for Nuclear Astrophysics, the Kavli Institute for Particle Astrophysics and Cosmology, the Korean Scientist Group, the Chinese Academy of Sciences (LAMOST), Los Alamos National Laboratory, the Max-Planck-Institute for Astronomy (MPIA), the Max-Planck-Institute for Astrophysics (MPA), New Mexico State University, Ohio State University, University of Pittsburgh, University of Portsmouth, Princeton University, the United States Naval Observatory, and the University of Washington.

 The Two Micron All Sky Survey was a joint project of the University of Massachusetts and the Infrared Processing and Analysis Center (California Institute of Technology). The University of Massachusetts was responsible for the overall management of the project, the observing facilities and the data acquisition. The Infrared Processing and Analysis Center was responsible for data processing, data
distribution and data archiving.

 This research has made use of NASA's Astrophysics Data System Bibliographic Services, the SIMBAD database, operated at CDS, Strasbourg, France, and the VizieR database of astronomical catalogues \citep{Ochsenbein2000}.  K.R.C gratefully acknowledges the support of NASA grant 80-0273 during the completion of this work.  Support for this work was provided by NASA through the Spitzer Space Telescope Fellowship Program.

\renewcommand{\thesection}{A\arabic{section}}
\setcounter{section}{0}  

\section{Appendix: Photometry of Parallax Standards on the SDSS system}\label{ap_1}

An accurate characterization of the color-magnitude diagram for 
low-mass stars is of critical importance for inferring 
stellar densities and distributions from purely photometric
datasets.  By mining the SDSS database, and conducting
supplementary observations with the NMSU 1m
telescope, we have obtained colors and magnitudes on the
SDSS photometric system for 76 stars and brown dwarfs with
measured trigonometric parallaxes.  

\subsection{Mining the SDSS database for Parallax Standards}

As the SDSS contains photometry of nearly a quarter of the
night sky, thousands of stars with measured trigonometric standards 
lie within the survey footprint.  While the vast majority of 
parallax standards are so bright that they saturate SDSS imaging,
parallax standards with servicable SDSS observations represent 
a valuable opportunity to map the main sequence color-magnitude 
relation directly from SDSS photometry, obviating the need to 
employ delicate, color-dependent transformations to place observations
from other telescopes onto the SDSS photometric system.  

To identify a catalog of parallax standards with native SDSS photometry, we
have searched the SDSS database for detections of objects contained within several 
publicly available parallax catalogs \citep{Monet1992,van-Altena1995,Tinney1995,Perryman1997,Dahn2002,Gould2004,Vrba2004}.
Culling the list to remove objects with flags indicative of severe 
photometric problems (ie, SATURATED-CENTER and/or INTERP-CENTER)
left 40 objects with native SDSS photometry and direct trigonometric
parallax measurements.  SDSS and 2MASS photometry for these objects, 
as well as their measured trigonometric parallaxes, are presented in Table
\ref{tab:nativephot}.  Figure \ref{jijCMR} displays the locations
of these sources in an $i-J$ vs. M$_J$ color-magnitude diagram.

\begin{deluxetable*}{lccccccccr}
   \tablewidth{0pt}
   \tabletypesize{\scriptsize}
   \tablecaption{Native SDSS Photometry of Parallax Standards   \label{tab:nativephot}}
   \tablehead{
     \colhead{Star} &
     \colhead{SDSS} &
     \colhead{SDSS} &
     \colhead{SDSS} &
     \colhead{2MASS} &
     \colhead{2MASS} &
     \colhead{2MASS} &
     \colhead{$\pi$} &
     \colhead{$\pi$} &
     \colhead{$\pi$} \\
     \colhead{Name} &
     \colhead{$i$} &
     \colhead{$r-i$} &
     \colhead{$i-z$} &
     \colhead{$J$} &
     \colhead{$J-H$} &
     \colhead{$H-K_s$} &
     \colhead{(mas/yr)} &
     \colhead{err.} &
     \colhead{ref.} }
\startdata
LHS 1970   & 16.26 &   0.83 &   0.49 &  14.58 &   0.58 &   0.12 &   12.9 &   0.7 & 1 \\
G 195-022  & 12.79 &   1.74 &   1.00 &  10.26 &   0.56 &   0.30 &   64.4 &   4.0 & 1 \\
LHS 2364   & 17.01 &   0.29 &   0.16 &  15.98 &   0.45 &  -0.03 &   38.3 &   2.7 & 1 \\
LP 323-239 & 13.55 &   1.66 &   0.90 &  11.18 &   0.44 &   0.26 &   48.1 &   1.9 & 1 \\
LHS 2828   & 14.81 &   2.23 &   1.14 &  12.00 &   0.52 &   0.33 &   36.5 &   0.9 & 1 \\
\\
LHS 2930   & 14.29 &   2.82 &   1.52 &  10.79 &   0.65 &   0.35 &  103.8 &   1.3 & 1 \\
LHS 3061   & 17.54 &   1.20 &   0.60 &  15.52 &   0.51 &   0.11 &   8.9 &   0.7 & 1 \\
J090551.11+553218.5 &   13.66 &   1.39 &   0.78 &  11.49 &   0.60 &   0.23 &   20.9 &   0.97 & 2 \\
J091841.52+582747.5 &  14.33 &   1.44 &   0.74 &  12.16 &   0.58 &   0.28 &   16.0 &   1.2 & 2 \\
J103327.6+472202.6 &  14.01 &   1.51 &   0.83 &  11.71 &   0.65 &   0.26 &   15.4 &   4.0 & 2 \\
\\
J113212.96+003632.5  & 14.18 &   1.44 &   0.95 &  11.78 &   0.52 &   0.29 &   21.2 &   4.0 & 2 \\
J113304.07+131816.9 & 14.91 &   1.36 &   0.69 &  12.79 &   0.61 &   0.27 &   10.8 &   1.5 & 2 \\
J120723.99+130213.9 & 15.68 &   2.01 &   1.05 &  13.00 &   0.60 &   0.32 &   24.1 &   1.4 & 2 \\
J123426.4+391309.4 &  14.21 &   1.36 &   0.76 &  12.10 &   0.57 &   0.24 &   14.4 &   3.1 & 2 \\
J132820.87+300319 & 16.73 &   2.50 &   1.37 &  13.31 &   0.67 &   0.35 &   20.7 &   2.0 & 2 \\
\\
J135921.6+251403.1 &  15.02 &   1.89 &   1.00 &  12.43 &   0.56 &   0.31 &   24.2 &   2.1 & 2 \\
J141436.71+160121.7 &  14.39 &   1.73 &   0.90 &  12.00 &   0.59 &   0.31 &   28.1 &   1.1 & 2 \\
J155719.67+074500.1 &  14.90 &   1.44 &   0.75 &  12.76 &   0.52 &   0.27 &   12.7 &   1.5 & 2 \\
LP374-4    &   17.01 &   0.29 &   0.16 &  15.98 &   0.45 &  -0.03 &   38.3 &   2.8 & 3 \\
LP130-226  &   15.32 &   2.54 &   1.28 &  12.20 &   0.51 &   0.33 &   39.6 &   1.1 & 3 \\
\\
LP323-239  &    13.55 &   1.66 &   0.90 &  11.18 &   0.44 &   0.26 &   49.0 &   2.2 & 3 \\
LP499-5    &   14.81 &   2.23 &   1.14 &  12.00 &   0.52 &   0.33 &   36.5 &   1.0 & 3 \\
LP98-79    &  14.29 &   2.82 &   1.52 &  10.79 &   0.65 &   0.35 &  10.38 &   1.4 & 3 \\
J003259.36+141036.4 &  23.09 &   1.16 &   3.69 &  16.83 &   1.18 &   0.70 &   30.1 &   5.2 & 4 \\
J010752.46+004156.3 &  21.19 &   2.02 &   2.55 &  15.82 &   1.31 &   0.80 &   64.1 &   4.5 & 4 \\
\\
J015141.68+124429.4 &  22.85 &   1.60 &   3.39 &  16.57 &   0.96 &   0.42 &   46.7 &   3.4 & 4 \\
TVLM 832-10443 &  17.08 &   2.86 &   1.67 &  13.13 &   0.69 &   0.48 &   36.0 &   0.4 & 5 \\
J074642.48+200031.6 &  16.08 &   2.57 &   1.85 &  11.76 &   0.75 &   0.54 &   81.9 &   0.3 & 5 \\
J082519.44+211550.3 &  20.61 &   2.80 &   2.76 &  15.10 &   1.31 &   0.76 &   93.8 &   1.0 & 5 \\
J083008.16+482847.2 &  21.22 &   2.35 &   3.14 &  15.44 &   1.10 &   0.67 &   76.4 &   3.4 & 4 \\
\\
TVLM 213-2005 &  17.14 &   2.77 &   1.66 &  13.39 &   0.65 &   0.48 &   30.1 &   0.4 & 5 \\
LHS 2471 &  14.70 &   2.60 &   1.50 &  11.26 &   0.60 &   0.40 &   70.2 &   2.1 & 5 \\
J125453.99-012247.4 &  22.25 &   2.05 &   4.22 &  14.89 &   0.80 &   0.25 &   75.7 &   2.9 & 4 \\
LHS 2924 &  16.19 &   2.39 &   1.80 &  11.99 &   0.76 &   0.48 &   95.0 &   5.7 & 5 \\
LHS 2930 &  14.29 &   2.82 &   1.52 &  10.79 &   0.65 &   0.35 &  103.3 &   1.3 & 5 \\
\\
J143517.28-004612.8 &  20.35 &   2.58 &   1.79 &  16.48 &   0.87 &   0.29 &   9.9 &   5.2 & 4 \\
J144600.48+002451.9 &  20.75 &   2.68 &   2.18 &  15.89 &   1.38 &   0.58 &   45.5 &   3.3 & 4 \\
TVLM 513-46546 &  16.08 &   2.64 &   1.79 &  11.87 &   0.69 &   0.47 &   94.4 &   0.6 & 5 \\
J171145.59+223204.2 &  22.15 &   1.38 &   2.33 &  17.09 &   1.29 &   1.07 &   33.1 &   4.8 & 4 \\
J225529.03-003434 &  19.85 &   2.36 &   1.91 &  15.65 &   0.89 &   0.32 &   16.2 &   2.6 & 4 \\
\enddata
\tablecomments{Reference 1: \citet{van-Altena1995} \\ Reference 2: \citet{Gould2004} \\ Reference 3: \citet{Monet1992} \\ Reference 4: \citet{Vrba2004} \\ Reference 5: \citet{Dahn2002} }
\end{deluxetable*}

\subsection{Observations of Parallax Standards with the NMSU 1m}

Additional constraints on the location of the lower main sequence
in SDSS/2MASS color-magnitude space can be derived by transforming
observations from secondary telescopes onto the native SDSS photometric
system.  Using time on the NMSU 1m telescope made available to the public 
through the NSF PREST program, we obtained $r'i'z'$ observations of 36 stars with measured 
trigonometric parallaxes, along with `deep' observations of fields containing 
late type stars with unsaturated SDSS photometry.  

The NMSU 1m telescope 
features a 2048x2048 E2V CCD providing a plate scale of 0.467 arcsec/pixel.
Observations were carried out in robotic mode under good conditions 
(clear to light cirrus, seeing 1-2\arcsec) on February 6th and 7th, 2007.  
Images were bias subtracted, flat fielded, and overscan corrected before 
instrumental magnitudes were measured via aperture photometry, using 
apertures with a 15 pixel radius and sky annuli 15 pixels in width 
with an inner radius of 25 pixels.  Observations of $u'g'r'i'z'$ photometric 
standards \citet{Smith2002} demonstrate internal photometric
accuracy of $\sigma$=0.04 magnitudes.

Comparing
instrumental magnitudes to SDSS database magnitudes for stars in the 
`deep' SDSS fields allow the derivation of airmass corrections simultaneously
with the zeropoints and color-terms required to transform from NMSU instrumental
magnitudes onto the SDSS photmetric system.  The accuracy of the NMSU to SDSS transformation
is limited by statistical uncertainties in NMSU detections of stars in the `deep' SDSS fields; residuals
in the transformed SDSS $i$ magnitudes range from 0.03 to 0.13 magnitudes for sources with $i=15-18$, respectively.  
Similar results are seen for stellar colors, with maximum $r-i$ and $i-z$ 
residuals at the faint ($i=18$) end of 0.15 and 0.2 magnitudes, respectively.
As a result, we adopt conservative uncertainties for our transformed SDSS photometry
of 0.15 magnitudes in $i$ and $r-i$, and 0.2 magnitudes in $i-z$.  

The SDSS magnitudes we derive for these sources from our transformed NMSU photometry 
are presented in Table \ref{tab:NMSUphot}, along with their 2MASS magnitudes and 
measured trigonometric parallaxes.  These sources are also shown in Fig. \ref{jijCMR}.

 \begin{deluxetable*}{lccccccccr}
   \tablewidth{0pt}
   \tabletypesize{\scriptsize}
   \tablecaption{NMSU-1m Photometry of Parallax Standards   \label{tab:NMSUphot}}
   \tablehead{
     \colhead{Star} &
     \colhead{SDSS} &
     \colhead{SDSS} &
     \colhead{SDSS} &
     \colhead{2MASS} &
     \colhead{2MASS} &
     \colhead{2MASS} &
     \colhead{$\pi$} &
     \colhead{$\pi$} &
     \colhead{$\pi$} \\
     \colhead{Name} &
     \colhead{$i$} &
     \colhead{$r-i$} &
     \colhead{$i-z$} &
     \colhead{$J$} &
     \colhead{$J-H$} &
     \colhead{$H-K_s$} &
     \colhead{(mas/yr)} &
     \colhead{err.} &
     \colhead{ref.} }
\}
\startdata
AC +61 23399             &  9.72 & 0.66 & 0.32 & 8.150 & 0.69 & 0.14 & 32.1 & 1.6 & 1 \\
BD+37 2337               & 9.29  & 0.29 &  0.12 & 8.033 & 0.56 & 0.11 & 24.99 & 1.6 & 1 \\
CN Leo                   & 10.41 & 2.39 & 1.38 & 7.085 & 0.60 & 0.40 & 425.0& 7.0 & 2 \\
G 88-28                  &  9.66 & 0.96 & 0.48 & 7.818 & 0.64 & 0.23 & 50.9 & 2.7 & 1 \\
G 96-1                   & 9.67  & 0.16 &  0.01 &  8.645 & 0.46 & 0.02 & 24.07 & 3.1 & 1 \\
\\
G 176-8                  &  9.76 & 0.76 & 0.41 & 8.107 & 0.67 & 0.22 & 30.1 & 2.3 & 1 \\
G 198-37                 & 9.23  & 0.69 &  0.33 &  7.610 & 0.64 & 0.20 & 43.25 & 2.0 & 1 \\
G 199-13                 &  9.09 & 0.22 & 0.06 & 8.000 & 0.50 & 0.10 & 22.7 & 1.2 & 1 \\
G 221-23                 & 8.37  & 0.47 &  0.20 &  6.873 & 0.61 & 0.15 & 57.53 & 2.7 & 1 \\
G 221-24                 & 9.23  & 0.95 &  0.45 &  7.379 & 0.58 & 0.23 & 57.53 & 2.7 & 1 \\
\\
G 222-1                  &  9.55 & 0.71 & 0.30 & 7.955 & 0.72 & 0.14 & 28.8 & 1.4 & 1 \\
G 250-34                 & 9.67  & 0.94 &  0.46 &  7.872 & 0.61 & 0.24 & 56.51 & 2.2 & 1 \\
G 256-10                 & 10.28 & 0.33 &  0.07 &  9.131 & 0.48 & 0.13 & 27.00 & 10.0 & 2 \\
GJ 3452                  &  9.68 & 1.07 & 0.54 & 7.772 & 0.55 & 0.27 & 76.3 & 2.6 & 1 \\
GJ 3504                  & 15.26 & 2.48 &  1.40 & 12.035 & 0.56 & 0.34 & 50.8  & 0.5 & 2 \\
\\
GJ 3855                  & 14.33 & 2.74 & 1.50 & 10.79 & 0.65 & 0.35 & 103.3& 1.3 & 2 \\
GJ 585.1                 & 9.15  & 0.53 &  0.23 &  7.693 & 0.61 & 0.18 & 39.0  & 1.7 & 1 \\ 
GJ 9455B                 & 10.01 & 0.19 & 0.04 & 8.948 & 0.38 & 0.13 & 20.7 &10.6 & 1 \\
GSC 03722-00302          & 9.70  & 0.30 &  0.09 &  8.450 & 0.51 & 0.12 & 20.39 & 2.0 & 1 \\
HD 24916                 & 7.31  & 0.34 &  0.12 & 6.063 & 0.58 & 0.15 & 63.41 & 2.0 & 1 \\
\\
HD 284552                & 9.78  & 0.38 &  0.14 &  8.427 & 0.59 & 0.15 & 24.98 & 2.0 & 1 \\
HIC 33805                &  9.72 & 0.64 & 0.32 & 8.129 & 0.59 & 0.19 & 43.9 & 2.2 & 1 \\
LHS 35                   & 10.68 & 1.59 & 0.80 & 8.424 & 0.50 & 0.27 & 143.0& 4.0 & 2 \\
LHS 197                  & 14.37 & 2.26 & 1.14  & 11.559 & 0.50 & 0.29 & 51.8 & 1.0 & 2 \\
LHS 224                  & 10.99 & 1.75 & 0.91 & 8.537 & 0.44 & 0.32 & 109.1 & 2.8 & 2 \\
\\
LHS 231                  & 14.01 & 1.78 & 0.90 &11.606 & 0.49 & 0.24 & 40.7 & 1.0 & 2 \\
LHS 287                  & 10.82 & 1.57 & 0.83  &  8.493 & 0.49 & 0.29 & 96.7  & 2.3 & 2 \\
LHS 303                  &  9.72 & 0.87 & 0.46 & 7.994 & 0.60 & 0.23 & 47.1 & 2.4 & 1 \\
LHS 331                  & 12.17 & 1.48 & 0.74  &  9.983 & 0.48 & 0.28 & 39.9  & 1.0 & 2 \\
LHS 2914                 & 12.04 & 0.29 & 0.13  & 10.803 & 0.47 & 0.16 & 27.0  & 13.0 & 2 \\
\\
LHS 1653                 & 10.35 & 0.45 & 0.22 & 8.926 & 0.61 & 0.13 & 28.7 & 3.4 & 1 \\
LTT 13340                &  9.29 & 0.32 & 0.15 & 8.036 & 0.53 & 0.12 & 23.2 & 1.8 & 1 \\
NLTT 36978               &  9.58 & 0.31 & 0.12 & 8.287 & 0.55 & 0.09 & 23.6 & 1.7 & 1 \\
NSV 5426                 & 10.14 & 0.35 & 0.15 & 8.835 & 0.55 & 0.15 & 23.0 & 2.2 & 1 \\
NSV 6424                 &  9.45 & 0.51 & 0.24 & 8.031 & 0.70 & 0.13 & 39.0 & 1.8 & 1 \\
\\
TVLM 263-71765           & 17.11 & 2.70 & 1.71 & 13.363  & 0.64 & 0.41 & 31.9 & 2.9 & 3 \\
\enddata
\tablecomments{Reference 1: \citet{Perryman1997} \\ Reference 2: \citet{van-Altena1995} \\ Reference 3: \citet{Tinney1995}}
\end{deluxetable*}



\setlength{\baselineskip}{0.6\baselineskip}






\end{document}